\documentclass[conference]{IEEEtran}
\IEEEoverridecommandlockouts
\usepackage{cite}
\usepackage{amsmath,amssymb,amsfonts}
\usepackage{algorithmic}
\usepackage{graphicx}
\usepackage{textcomp}
\usepackage{xcolor}
\usepackage{graphicx}
\usepackage{textcomp}
\usepackage{wrapfig}
\usepackage{xspace}
\usepackage{subfig}
\usepackage{soul}
\usepackage{url}

\newcommand{\trefi}{$T_{REFP}$\xspace}
\newcommand{\trefp}{$T_{REFP}$\xspace}
\newcommand{\vdd}{$V_{DD}$\xspace}
\newcommand{\remove}[1]{}

\def\BibTeX{{\rm B\kern-.05em{\sc i\kern-.025em b}\kern-.08em
    T\kern-.1667em\lower.7ex\hbox{E}\kern-.125emX}}
\begin{document}

\title{Workload-Aware DRAM Error Prediction using Machine Learning}

\author{\IEEEauthorblockN{Lev Mukhanov\IEEEauthorrefmark{1}, Konstantinos Tovletoglou\IEEEauthorrefmark{2}, Hans Vandierendonck\IEEEauthorrefmark{3}, \\Dimitrios S. Nikolopoulos\IEEEauthorrefmark{4}\textbf{(*)} and Georgios Karakonstantis\IEEEauthorrefmark{5}}
\IEEEauthorblockA{School Of Electronics, Electrical Engineering And Computer Science, ECIT,
Queen's University Belfast, UK\\
Email: \IEEEauthorrefmark{1}l.mukhanov@qub.ac.uk,
\IEEEauthorrefmark{2}ktovletoglou01@qub.ac.uk,
\IEEEauthorrefmark{3}h.vandierendonck@qub.ac.uk,\\
\IEEEauthorrefmark{4}d.nikolopoulos@qub.ac.uk
\IEEEauthorrefmark{5}g.karakonstantis@qub.ac.uk}
\\
\IEEEauthorblockA{\IEEEauthorrefmark{4}\textbf{(*)} The present affiliation: Department of Computer Science, Virginia Polytechnic Institute and State University, USA\\} 
}
  
\maketitle
\begin{abstract}
The aggressive scaling of technology may have helped to meet the growing demand for higher memory capacity and density, but has also made DRAM cells more prone to errors. 
Such a reality triggered a lot of interest in modeling DRAM behavior for either predicting the errors in advance or for adjusting DRAM circuit parameters to achieve a better trade-off between energy efficiency and reliability. Existing modeling efforts may have studied the impact of few operating parameters and temperature on DRAM reliability using custom FPGAs setups, however they neglected the combined effect of workload-specific features that can be systematically investigated only on a real system.

In this paper, we present the results of our study on workload-dependent DRAM error behavior within a real server considering various operating parameters, such as the refresh rate, voltage and temperature. We show that the rate of single- and multi-bit errors may vary across workloads by 8x, indicating that program inherent features can affect DRAM reliability significantly. Based on this observation, we extract 249 features, such as the memory access rate, the rate of cache misses, the memory reuse time and data entropy, from various compute-intensive, caching and analytics benchmarks. We apply several supervised learning methods to construct the DRAM error behavior model for 72 server-grade DRAM chips using the memory operating parameters and extracted program inherent features.
Our results show that, with an appropriate choice of  program  features  and  supervised learning method, the rate of single- and multi-bit errors can be predicted for a specific DRAM module with an average error of less than 10.5~\%, as opposed to the 2.9x estimation error obtained for a conventional workload-unaware error model. 
Our model enables designers to predict DRAM errors in advance for less than a second and study the impact of any workload and applied software optimizations on DRAM reliability.
\end{abstract}

\section{Introduction}

The worsening of parametric variations in deep nanometer technologies and aggressive scaling of circuit parameters for low power operation made memory cells more prone to errors, the number of which may vary significantly across different 
The manifestation of such errors, that depends on various
factors~\cite{Liu:2013:ESD:2485922.2485928,5766061,6043592,7092438,Tsiokanos:2018:VPC:3218603.3218617,8474084,8714942} related to circuit parameters, temperature, as well as system architecture, and workloads,
threaten the availability of computing systems and quality of service of sensitive storage components in data centers~\cite{Schroeder:2009:DEW:1555349.1555372}
and supercomputers~\cite{7877133, Sridharan:2012:SDF:2388996.2389100, Hwang:2012:CRD:2150976.2150989}.
The increased risks have triggered few research studies on prediction of DRAM errors in advance~\cite{Giurgiu:2017:PDR:3154448.3154451,Lan:2010:SDM:1786811.1787078,1633531,Sahoo:2003:CEP:956750.956799,Yu:2011:POF:2056318.2057092,7575388,DBLP:conf/dsn/NieXGPEST18}.
However, these studies were performed only for DRAM operating under nominal circuit parameters and typical environmental conditions. Moreover, even though they tried to consider other workload/architecture related factors, this was limited due to the constrained access to only specific features, like percentage of utilized memory, average CPU utilization and hardware characteristics~\cite{Meza:2015:RME:2859844.2859952}. The joint consideration of more features may reveal new non-linear behaviors that cannot be captured by linear regression models~\cite{Meza:2015:RME:2859844.2859952} or traditional workload-agnostic statistical models~\cite{5165089}.
In addition, all these studies lacked an adequate number of samples because of the rare manifestation of errors for DRAM operating under nominal circuit parameters, which may result in contradictory observations~\cite{Schroeder:2009:DEW:1555349.1555372,Meza:2015:RME:2859844.2859952}.

In the past, there have been several experimental studies that tried to predict the error behavior of DRAM operating under non-nominal circuit parameters~\cite{Liu:2013:ESD:2485922.2485928}, such as the refresh period (\trefp) and the supply voltage (\vdd),
and even under various temperatures~\cite{678551,Khan:2014:EEM:2637364.2592000,Liu:2013:ESD:2485922.2485928,8474184,Mukhanov:2018:CHW:3229631.3236091}. However, the main goal of these studies was to improve DRAM performance and energy efficiency by scaling \trefp or \vdd \cite{8342175,8416198}, rather than model DRAM errors.
Although, some of these works have indicated the fact that the certain program features, such as the pattern of data stored in memory\cite{Liu:2012:RRI:2337159.2337161, Venkatesan:2006:RAPID, Raha:2015:QDA:2830689.2830702, 6579591, Patel:2017:RPE:3140659.3080242, Wang:2015:RCR:2744769.2744897,wang2018reducing,8046197,wang2018reducing, refrint, 4408251}, may change the number of manifested errors, none of them attempted to jointly consider the impact of DRAM circuit parameters and various program inherent features on DRAM reliability. Beside the data pattern, program inherent features encapsulate features that can be extracted using hardware program counters, e.g. the processor utilization, the rate of memory and cache misses, $IPC$. The program counters may have been used in the past for power and performance modelling\cite{854380,Mukhanov:2017:AFE:3058793.3050436,7429297}, but were never used for DRAM reliability modeling in conjunction with various circuit parameters and temperature. 
Modeling the joint impact of such a wide range of features requires a novel experimental framework implemented on a real system with a complete software stack. This framework, unlike the custom FPGA setups used in prior studies~\cite{Liu:2012:RRI:2337159.2337161, Raha:2015:QDA:2830689.2830702, DBLP:journals/dt/JungMRWW17}, should be capable of running real workloads under different DRAM temperatures and provide a mechanism to measure errors and hardware program counters.

\textbf{The main goal} of this work is to systematically investigate the effect of various program inherent features on DRAM reliability and develop a DRAM error model that takes into consideration the combined effect of these features, as well as the reliability variation across chips, DRAM circuit parameters and temperature. This model enables designers to predict DRAM errors based on few workload-specific features for a given set of DRAM circuit parameters and temperature. Such a prediction does not require long-running DRAM characterization campaigns that may take hours or even days on complex experimental setups. The error behavioral model facilitates: i) evaluating how prone to errors are specific workloads; ii) evaluating the implicit impact of applied software optimizations (e.g. compiler, or thread level parallelism) on DRAM reliability; iii) predicting maintenance cycles, as aimed by recent works~\cite{Hwang:2012:CRD:2150976.2150989, Meza:2015:RME:2859844.2859952}; iv) guiding the adjustment of the circuit DRAM parameters for saving energy~\cite{Liu:2011:FSD:1961296.1950391, Raha:2015:QDA:2830689.2830702}.

Our contributions can be summarized as follows:
\begin{itemize}
\item We develop a novel experimental framework for characterizing DRAMs under relaxed refresh period and lowered supply voltage within a state-of-the-art 64-bit ARM based server. In order to experiment under different DRAM temperatures, we implement a thermal testbed that allows us to fine tune the temperature of each DIMM on the server.

\item We perform a characterization of 72 server-grade DRAM chips under scaled refresh period and lowered supply voltage running compute-intensive, caching, and analytics benchmarks. Our study shows that the rate of single- and multi-
bit errors may vary across workloads and DRAM chips by 8$\times$ and 188$\times$, respectively. 

\item To quantify the dynamically changing data and access patterns of a running program, we introduce new metrics, namely the DRAM reuse time and the data entropy. We extract these program features along with 247 features measured using hardware performance counters during the execution of each workload and correlate them with DRAM errors, identifying features that are more likely to affect DRAM reliability.

\item We apply three different Machine Learning methods to train a workload-aware DRAM error model based on the extracted program features, DRAM circuit parameters and temperature. In particular, we investigate the accuracy of Support Vector Machines (SVM), K-nearest neighbors (KNN) and Random Decision Forests (RDF). We compare these models on 4 different DRAM devices considering various sets of program features used for training. Our study shows that the highest accuracy of DRAM error estimates is achieved by KNN, which enables us to predict DRAM error rates within 300~ms
with an average error that does not exceed 10.2~\%, as opposed to the 2.9$\times$ estimation error obtained for a conventional workload-unaware error model. We make the DRAM error behavioral model (KNN-based) publicly available, which will be periodically updated based on new characterization results~\cite{dwer}.
\end{itemize}
\section{Background}\label{background}
\begin{figure}[t]
        \centering
        \includegraphics[width=\columnwidth, keepaspectratio]{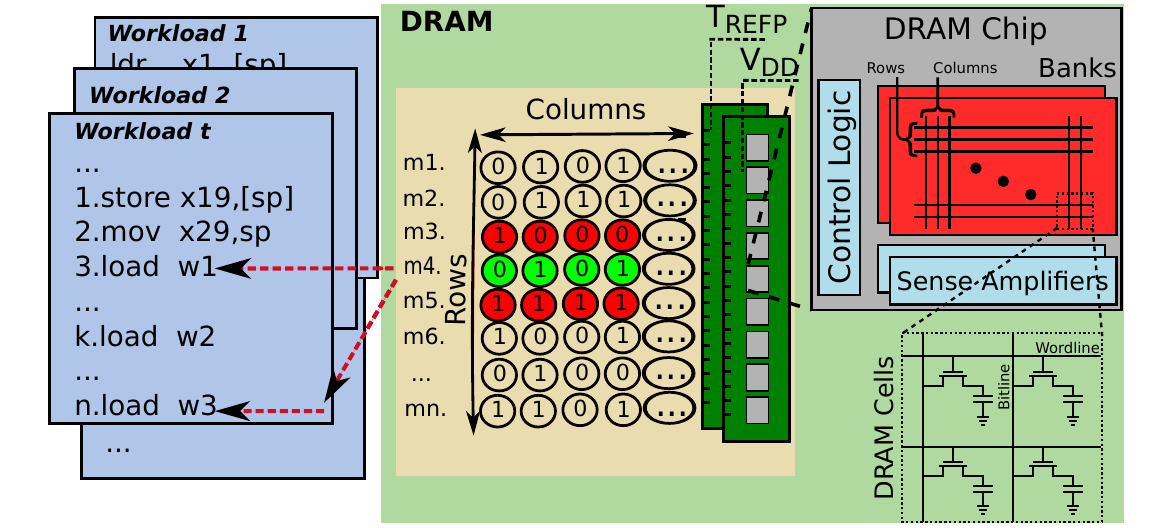}
        \caption{Interaction between workloads and DRAM; Internal structure of DRAM.}
        \label{fig:apm_xgene}
\end{figure}

\subsection{DRAM Basics}
DRAM is an essential component in any modern computing system, used to realize the memory subsystem. Beside the data caches, the memory subsystem includes several channels (Memory Channel Units, MCUs) which are used to transfer data and commands between the processor and DRAM. Each channel is connected to a number of Dual In-line Memory Modules (DIMMs).
A DIMM usually has two ranks that contain DRAM chips. Within each chip, DRAM cells are organized into banks, which are two-dimensional arrays that can be accessed in parallel based on rows and columns (see Figure~\ref{fig:apm_xgene} on the right). The basic storage element of a DIMM is a cell, consisting of a transistor and a capacitor. When a row of cells is accessed, the peripheral circuitry of a DIMM senses the data stored in this row via amplifiers and sends it to the processor.

\subsection{DRAM Error Behavior: Main Operating Parameters}
The main drawback of the DRAM technology is the limited \emph{retention time}~\cite{Liu:2013:ESD:2485922.2485928} of a cell's charge. To avoid any error induced by the charge leakage, DRAM employs an \emph{Auto-Refresh} mechanism that recharges the cells in the array periodically~\cite{Liu:2013:ESD:2485922.2485928}. Conventionally, all DDR technologies adopt a \emph{refresh period}, \trefp, of 64 $ms$ for refreshing each cell. Other critical parameter that affects DRAMs' power and reliability behavior is the supply \emph{voltage}, \vdd. Similar to \trefp, \vdd of DRAM chips is chosen conservatively by vendors to ensure that each chip operates correctly under a wide range of conditions. 
In addition to the above circuit parameters, one of the main environmental conditions that affect DRAM reliability is temperature ($TEMP_{DRAM}$). In fact, it has been reported that the retention time of DRAM cells decreases exponentially with increasing temperature~\cite{678551}.

\subsection{DRAM Error Behavior: Workload-Dependent Parameters}
The use of DRAM depends on executed instructions that access the memory in a certain way. In particular, the data read and written by a program (\textit{data pattern}) from/to memory and the order in which the program refers to this data (\textit{access pattern}) vary across workloads. Note that access pattern also encapsulates the rate of memory accesses and the average time between accesses to DRAM cells. 
\begin{wrapfigure}{R}{0.5\columnwidth}
\centering
\includegraphics[width=0.49\columnwidth, keepaspectratio]{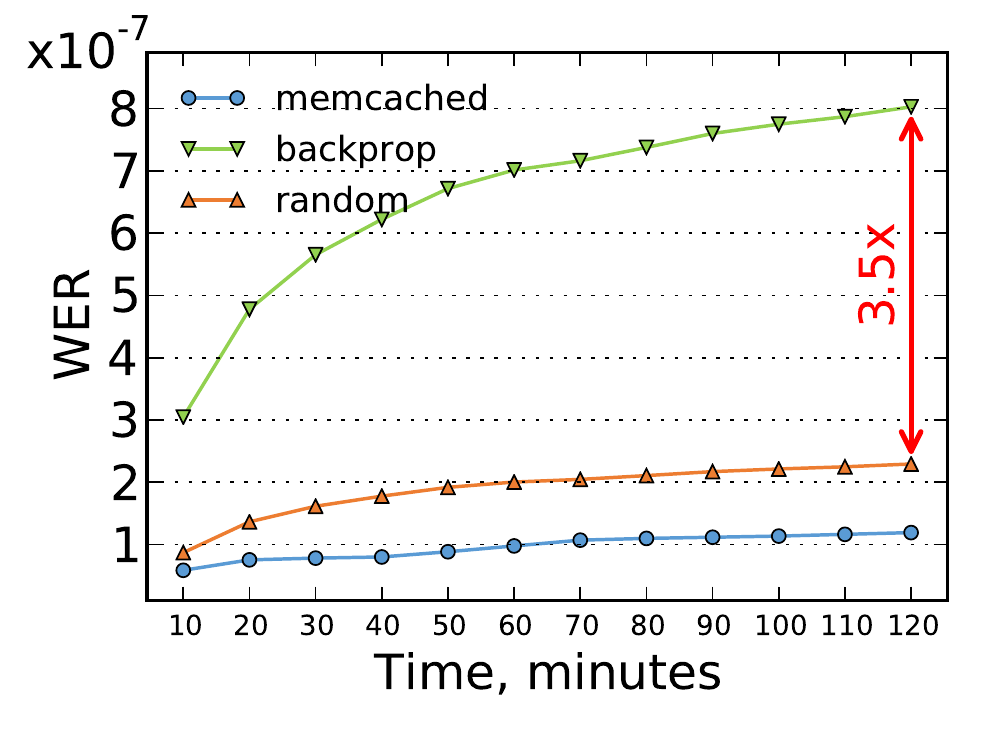}
\caption{The rate of single-bit errors per 64-bit word (WER) when running \emph{memcached}, \emph{backprop} and the \emph{random} micro-benchmark for DRAM operating under $2.283~s$ \trefp and lowered \vdd ($1.428~V$) at $70^{\circ}C$ (2 hours run, 8 threads).}
\label{fig:error_dif_example}
\end{wrapfigure}

Previous studies have demonstrated that the data pattern of a running program may affect DRAM errors~\cite{Liu:2013:ESD:2485922.2485928,Khan:2014:EEM:2637364.2592000}. Meanwhile, the frequency of read and write accesses (i.e. the memory access pattern) may reduce the number of manifested errors, since each read/write naturally refreshes DRAM~\cite{refrint,8046197}. We demonstrate such accesses in Figure~\ref{fig:apm_xgene} where $3.load$ and $n.load$ instructions from the $t$-th workload refresh data in the $m4$ DRAM line. 
By contrast, if a row is accessed many times, then some cells from neighbouring rows may leak charge due to the DRAM cell-to-cell interference~\cite{6853210}. This effect has been exploited widely for "row hammer" attacks~\cite{vanderVeen:2016:DDR:2976749.2978406,Mutlu:2017:RPO:3130379.3130643}. Specifically, data in the $m3$ and $m5$ DRAM rows (see Figure~\ref{fig:apm_xgene}) may be compromised when the $m4$ row is accessed too often. Thus, by increasing the memory access frequency to the same row, we reduce the number of errors manifested in this particular row, while inducing errors in neighborhood rows due to the DRAM cell-to-cell interference. Accordingly, inherent program features that change memory data and access patterns of a running workload may have an important effect on DRAM reliability. However, to the best of our knowledge, none of the previous studies have systematically investigated the combined effect of data and access pattern on DRAM reliability under relaxed DRAM parameters and varying DIMM temperature. 

Failing to identify the combined effect of program features
on real server deployments may limit or nullify the efficacy of existing approaches. For example, several previous studies have proposed fine-grained methods to control DRAM parameters based on the retention time measured for each cell~\cite{Liu:2012:RRI:2337159.2337161,7266870}. To measure the retention time, authors use micro-benchmarks that implement the worst-case data pattern manifesting errors in the vast majority of error-prone memory locations~\cite{Khan:2014:EEM:2637364.2592000,6579591,DBLP:journals/dt/JungMRWW17,678551}. However, our study shows that real applications may trigger errors in many more memory locations than the conventional data pattern micro-benchmarks. Figure~\ref{fig:error_dif_example} depicts the rate of single-bit errors per 64-bit word ($WER$) observed for DRAM operating under relaxed parameters when running two different benchmarks (\emph{memcached} and \emph{backprop}), and the most stressful data pattern micro-benchmark (the \emph{random} data pattern micro-benchmark~\cite{Liu:2013:ESD:2485922.2485928}). We see that the $WER$ incurred by \emph{backprop} is 3.5$\times$ higher than the rate observed for \emph{random}. As a result, the cell retention time measured using this data pattern micro-benchmark may be inaccurate, which, in turn, may lead to uncertain hardware behavior or even hardware crashes when the proposed methods are applied in practice. On the other hand, the proposed methods may be too pessimistic about the retention time and thus ineffective, since real applications, such as \emph{memcached}, may trigger errors in fewer memory locations than the micro-benchmark. These results indicate that designers should take into account the combined effect of workload-dependent factors on DRAM reliability when designing error mitigation techniques.

\remove{
\begin{wrapfigure}{R}{0.5\columnwidth}
\centering
\includegraphics[width=0.49\columnwidth, keepaspectratio]{figures/error_dif_example.pdf}
\caption{The number of single-bit errors detected by ECC and IPC  measured for \emph{kmeans} and \emph{memcached} for DRAM operating under $2.283~s$ \trefp and lowered \vdd ($1.428~V$) at $50^{\circ}C$ (2 hours run, 8 threads)}
\label{fig:error_dif_example}
\end{wrapfigure}

To investigate the effect of workloads on DRAM reliability, we run an HPC benchmark (\emph{kmeans}) and a real Cloud workload (\emph{memcached}) when DRAM operates under relaxed parameters (the refresh rate increased by 35x and the DRAM voltage reduced by 5\%) at $50^{\circ}C$. Figure~\ref{fig:error_dif_example} shows the number of single-bit errors detected by ECC when running two benchmarks. We see that the number of single-bit errors differs by $45\times$ for \emph{kmeans} and \emph{memcached}, even though both applications allocate the same size of memory. Such a difference can be explained by workload inherent features. For example, the Instruction Per Clock (IPC) of \emph{kmeans} is 1.5x higher than that measured for \emph{memcached} (see Figure~\ref{fig:error_dif_example}). 

The above indicates that there are could be program inherent features affecting DRAM error behavior that have not been discovered previously. In this study, we investigate for the first time systematically and quantify the correlation between various program features, including architecture-dependent features, and DRAM errors. 
}
\subsection{DIMM-to-DIMM Variation}
Apart from the above circuit and workload-dependent parameters, DRAM reliability may vary across DIMMs from different vendors~\cite{Khan:2017:DMD:3123939.3123945,Liu:2013:ESD:2485922.2485928}, and even across DIMMs manufactured by the same vendor. This variation is due the manufacturing process~\cite{5165089} and the internal design of DRAM modules, such as true-anti cell organization~\cite{Liu:2013:ESD:2485922.2485928}, address scrambling~\cite{994601,Khan:2017:DMD:3123939.3123945} and the remapping of faulty cells~\cite{7579745}. Our study indicates that the rate of single-bit errors per 64-bit word may vary by 188$\times$ across different DRAM chips.

\subsection{Challenges}
According to the above discussion, there are various cross-layer parameters, at the circuit (e.g. \vdd, \trefp), micro-architecture (i.e. cache organization and DRAM architecture), application (i.e. data and DRAM access patterns) layers, which in combination with environmental parameters (i.e. the DRAM temperature) can significantly influence DRAM reliability.
Predicting the potential failures early at design or operation cycle by considering all the combined cross-layer effects  is an extremely challenging problem.
\section{DRAM Error Prediction}
\label{sec:modeling}
\subsection{Mathematical Formulation of the Problem}
Let us assume that a workload, having a specific set of program features ($Ftrs=(f_{1},f_{2},...,f_{K})$ where $f_{i}$ is the $i$-th feature), allocates data on a DRAM device ($Dev$) when this device operates under \trefp and \vdd at a certain temperature ($TEMP_{DRAM}$). Then, to predict a target DRAM error metric $M_{err}$ for this workload, we need to model a prediction function ($M$) such that:
\begin{equation}
\footnotesize
\begin{aligned}
 M_{err}=M(Ftrs,Dev,T_{REFP},V_{DD},TEMP_{DRAM})
\end{aligned}
\label{math:statement}
\end{equation}

It is evident that building such a model is extremely challenging due to the number of possible parameter combinations. To address this challenge, we propose to use a supervised Machine Learning (ML) technique, since we believe it is hard to find an analytical model that predicts DRAM error behavior accurately considering the DIMM-to-DIMM variation and all the parameters.  
\subsection{ML Models}
\label{subsec:models}
In our study, we investigate the accuracy of the following Machine Learning models: Support Vector Machines (SVM),  K-nearest neighbors algorithm (KNN) and  Random Decision Forests (RDF).
These models have a high accuracy  for both linear and non-linear prediction problems~\cite{Fernandez-Delgado:2014:WNH:2627435.2697065}. We use the scikit-library to implement the models ~\cite{scikit}.

\subsection{DRAM Error Metrics}
\label{section:experimental:dynamorio}
\begin{table}
\centering
    \begin{tabular}{ | c | c | c | }
    \hline
    \textbf{Num. of corrupted bits} & \textbf{Type of errors} & \textbf{Abbreviation}\\ \hline
    1 & corrected & CE  \\ 
    \hline
    $> 1$ & uncorrected/detected & UE  \\ 
    \hline
    $> 2$ & uncorrected/undetected & SDC\\
    \hline
    \end{tabular}
    
    \caption{Types of DRAM errors that can be corrected or detected with ECC SECDED.}
    \label{ecc_secded}
\end{table}

There are several types of errors that may manifest in DRAM chips~\cite{5933,1029773,1299234,Sridharan:2015:MEM:2786763.2694348}. Vendors implement a special hardware (ECC, Error Correction Codes) in server-grade chips to automatically correct such errors.  In this study, we use hardware that supports ECC SECDED, which is implemented in the majority of commercial servers. There are three types of memory errors that may occur when ECC SECDED is enabled (see Table~\ref{ecc_secded}): single-bit errors (or correctable errors, \textbf{CE}); detected errors where more than one bit in a 64-bit word is corrupted (or uncorrectable errors, \textbf{UE}); and errors where more than 2 bits are corrupted per word, which are not corrected and not detected by ECC. The last types of errors manifest so-called Silent Data Corruption (\textbf{SDC}), since such errors are invisible for hardware.

\textbf{Correctable errors:} To characterize DRAM in terms of CEs, we measure the rate of single-bit errors per 64-bit, $WER$, for the amount of memory used by an application as:
\begin{equation}
WER=\frac{N_{CE}}{MEM_{SIZE}} 
\label{wer_estimate}
\end{equation}
where $N_{CE}$ is the number of unique 64-bit word locations where CEs have manifested and $MEM_{SIZE}$ is the size (in 64-bit words) of memory allocated by the application. $WER$ shows the probability of a word being erroneous regardless of the size of memory allocated by the application.

\textbf{Uncorrectable Errors:} To characterize DRAM in terms of UEs, we estimate the probability of an UE, triggered by a running application as:
\begin{equation}
P_{UE}=\frac{N_{UE}}{N_{EXP}}
\label{pcrash_estimate}
\end{equation}
where $N_{UE}$ is the number of experiments with the application that resulted in an UE, and $N_{EXP}$ is the total number of experiments with the application. 

\subsection{Program Inherent Features}
To investigate software-level factors that may affect DRAM reliability, we extract the following program features.

\textbf{The DRAM Reuse Time:} 
The DRAM reuse time ($T_{reuse}$) is the average time between memory accesses to the same 64-bit word (or a DRAM location). This metric is important for our study, as memory accesses inherently refresh the stored charge~\cite{refrint,8046197}, while $T_{reuse}$ denotes the average period between accesses to the DRAM cells, and thus, the average refresh period of cells incurred by memory accesses. If $T_{reuse}<T_{REFP}$ for a running program, then the number of DRAM errors induced by the charge leakage will decrease. We estimate $T_{reuse}$ by averaging the DRAM reuse time over all memory accesses, i.e., $T_{reuse} = \frac{\sum_{i=1}^{N_{mem}} T^{i}_{reuse}}{N_{mem}}$, where $T^{i}_{reuse}$ is the reuse time for the $i$ memory access instruction with reference to some address. In turn, we calculate $T^{i}_{reuse}$ as:
\begin{equation}
T^{i}_{reuse}=CPI \times D^{i}_{reuse}
\end{equation}
In this equation, $CPI$ is the average number of clock cycles per instruction measured for an entire program, $D^{i}_{reuse}$ is the number of instructions executed since the last reference to the address accessed by the $i$ instruction. We extract $D^{i}_{reuse}$ using a dynamic binary instrumentation tool, DynamoRIO~\cite{2011:TBD:2190025.2190038}.\remove{Specifically, we run each benchmark with DynamoRio using the following options: \texttt{-t drcachesim -simulator\_type reuse\_time -page\_size 64K}.} We validated $T_{reuse}$ estimates using micro-benchmarks where we can control and measure $T_{reuse}$ for specific memory accesses, and found that the approximation is accurate.

\begin{figure*}[htp]
        \begin{minipage}[t]{0.64\textwidth}
                \centering
                \includegraphics[width=\columnwidth]{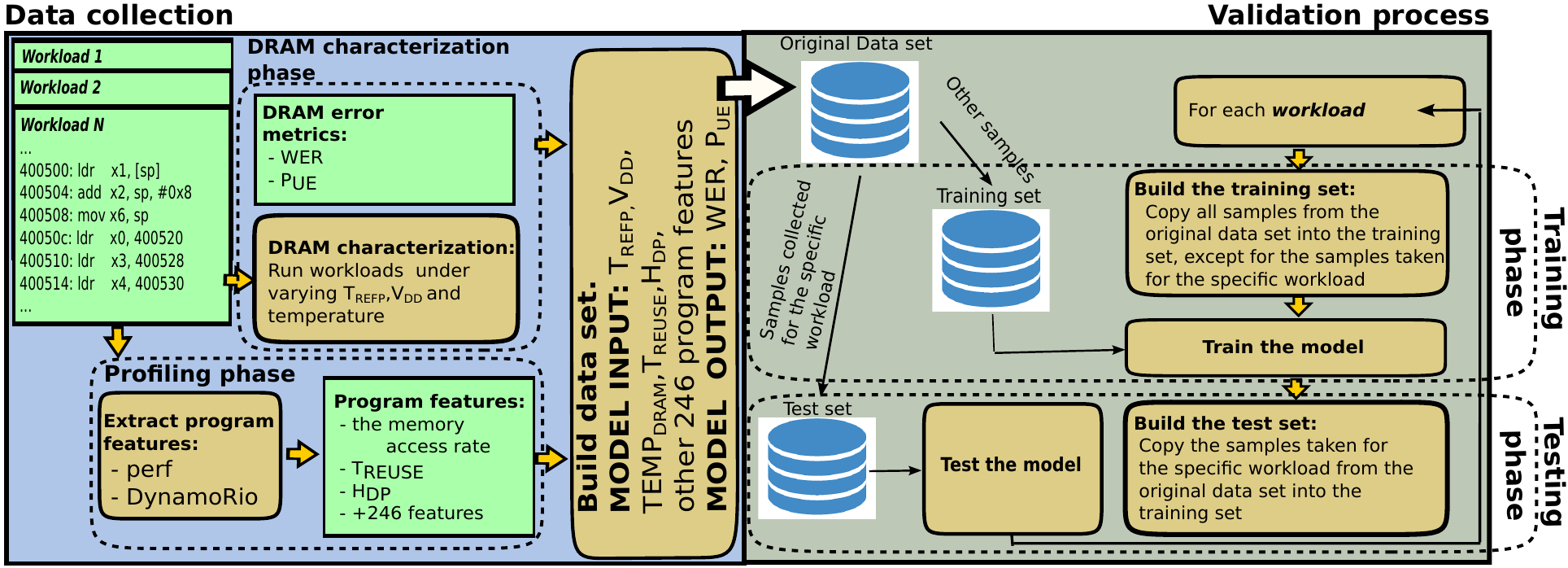}
                \caption{Overview of the data collection and validation processes.}
               \label{fig:training}
        \end{minipage}%
        \hfill%
        \begin{minipage}[t]{0.35\textwidth}
                \centering
                \includegraphics[width=\columnwidth]{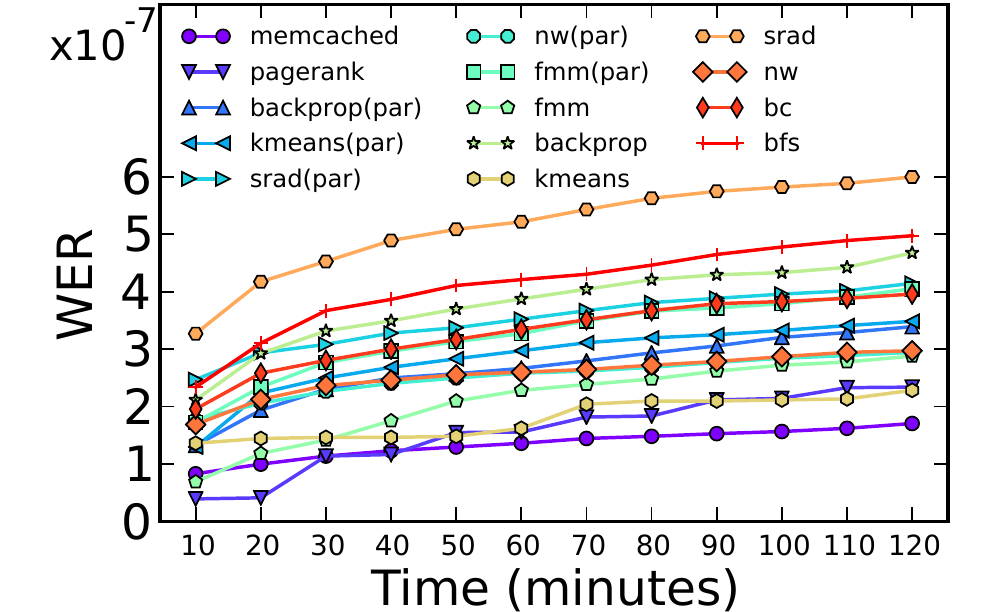}
                \caption{$WER$ of each benchmark under $2.283~s$ \trefp and $1.428~V$ \vdd at $50^{\circ}C$.}
                \label{fig:wer_in_time}
        \end{minipage}
\end{figure*}

\textbf{The Data Entropy:} 
To quantify the varying data patterns (DPs) stored in memory across different time instances, we introduce a new metric, the DP entropy, $H_{DP}$. To estimate $H_{DP}$, we profile all workloads with DynamoRIO and take samples of the data for each write memory access that is ultimately stored in DRAM. We then estimate $H_{DP}$ based on the sampled data as: 
\begin{equation}
H_{DP} = -\sum_{i=0}^{2^{32}-1}P(x_i)\times log_2(P(x_i));
P(x_i) = \frac{N_{WR}(x_i)}{N_{WR}^{TOT}}
\end{equation}
where $N_{WR}(x_i)$ is the number of writes operations with data $x_i$ in a word and $N_{WR}^{TOT}$ is the total number of writes.

\textbf{Performance Counters:} Another important parameter that may affect DRAM reliability is the number of memory accesses executed per cycle, as the cell-to-cell interference grows with the rate of memory accesses~\cite{5727538,1029773}. We measure this number, along with 247 program metrics, such as L1/L2/memory accesses (writes and reads) per cycle, and IPC and the SoC utilization, using existing hardware performance counters (\emph{perf}) to investigate the potential effect of other architecture-level parameters on DRAM error behavior.

\subsection{Data Collection} 
To collect data for training of the ML models, we run a set of representative benchmarks (workloads) under varying DRAM operating parameters, such as \trefp, \vdd and temperature, and measure $WER$ and $P_{UE}$, as shown in Figure~\ref{fig:training}. We additionally run each benchmark to collect all the inherent program features using DynamoRio and the \emph{perf} tool (\textit{Profiling phase}). Then, we combine collected program inherent features with the $WER$ or $P_{UE}$ measurements. 

\subsection{Accuracy Evaluation of ML Models}We evaluate accuracy of the ML models using the cross-validation technique~\cite{Kohavi:1995:SCB:1643031.1643047} by partitioning the collected data into a test set and a training set. We use the Leave-One-Out~\cite{Mukherjee2006} partitioning as shown in Figure~\ref{fig:training}. In particular, for each benchmark we create a test set that consists of samples taken only for a specific benchmark, whereas the training set contains all other samples. We train the model (\textit{Training phase}) and test (\textit{Testing phase}) its prediction accuracy for each pair of training and testing sets (see Figure~\ref{fig:training}). Finally, we average the prediction accuracy over all testing experiments, the number of which is equivalent to the total number of benchmarks.
\section{Experimental Setup}\label{sec:framework}

\begin{figure}[t]
        \begin{minipage}[t]{0.48\linewidth}
                \includegraphics[width=\columnwidth]{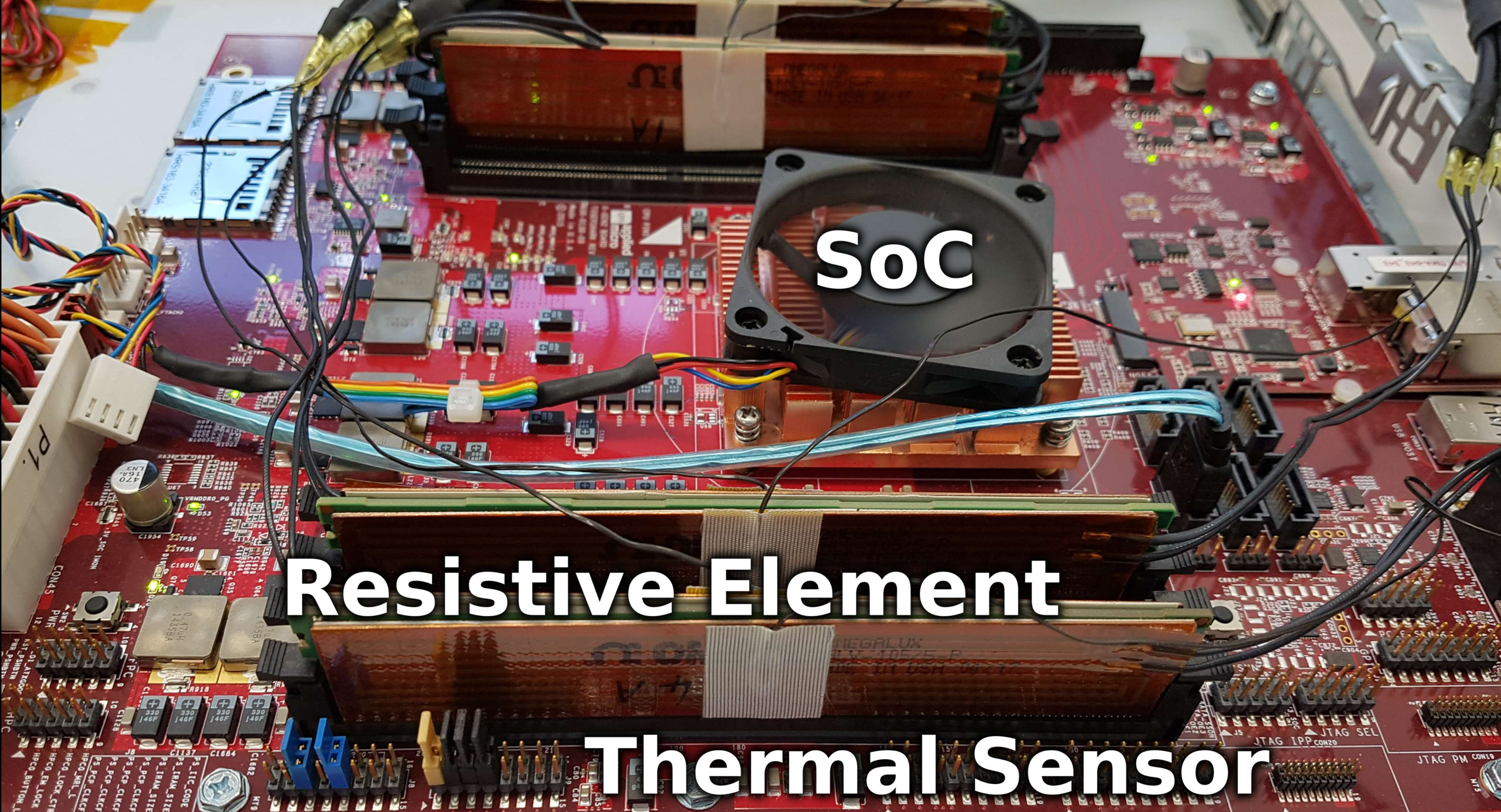}
                \caption{X-Gene2 with the custom thermal adapters.}
                \label{fig:thermal:a}
                \label{fig:bit_distribution}
        \end{minipage}%
        \hfill%
        \begin{minipage}[t]{0.48\linewidth}
                \includegraphics[width=\columnwidth]{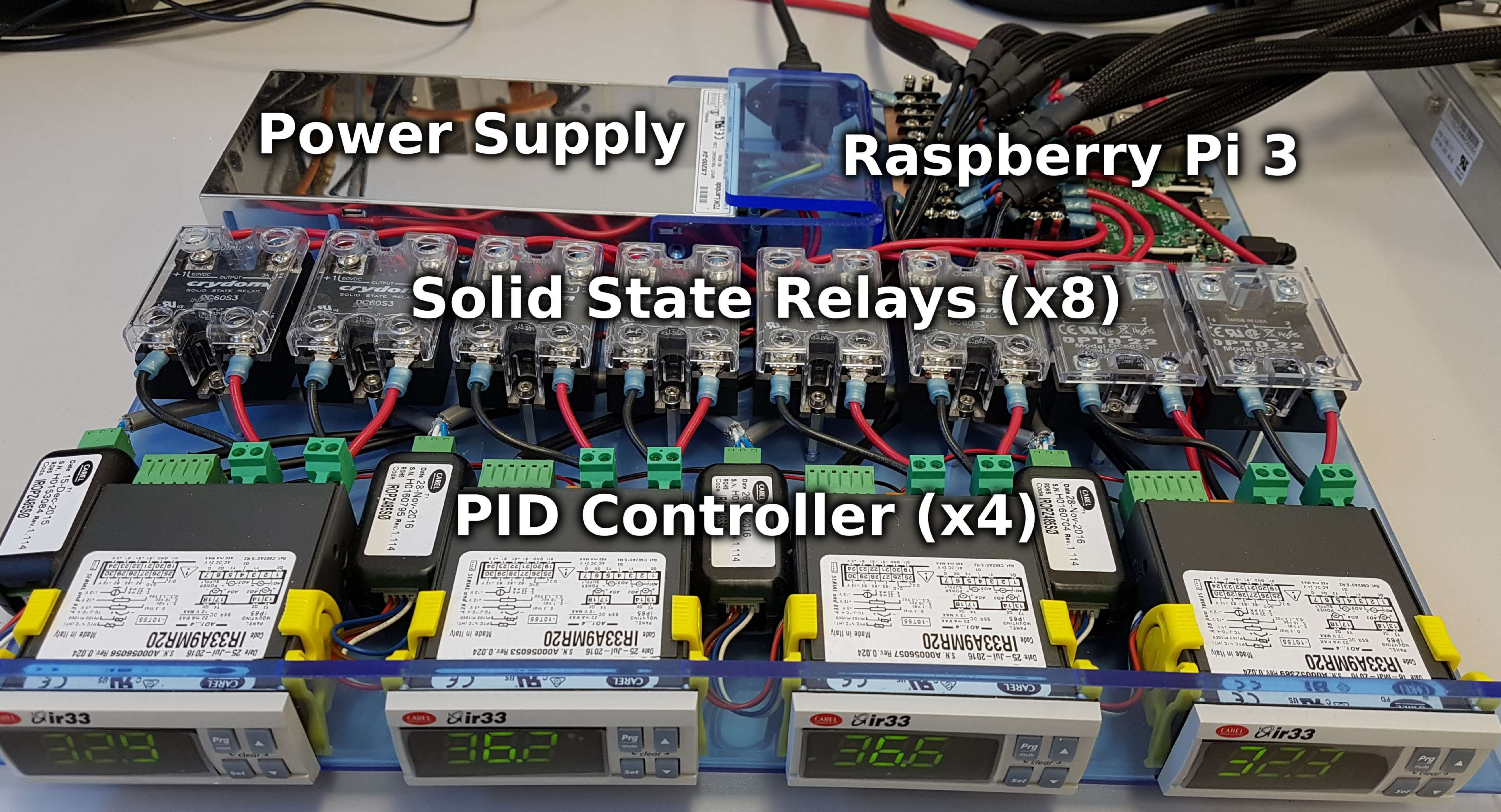}
                \caption{Temperature controller board.}
                \label{fig:thermal:b}
        \end{minipage}
\end{figure}
To enable DRAM characterization, we developed a unique experimental setup which we discuss in this section. 

\subsection{Experimental Framework}
The basis of our experimental framework is a state-of-the-art commodity 64-bit ARMv8-based server, the X-Gene2 Server-on-a-Chip. The X-Gene2 SoC consists of eight 64-bit ARMv8 cores running at 2.4GHz. The X-Gene2 has four DDR3 Memory Controller Units (MCUs). 
In our campaign, we are experimenting with 4 Micron DDR3 8GB DIMMs at 1866~MHz~\cite{micron-datasheet}, with one DIMM per MCU. In total, we are characterizing 72 chips of 4Gb x8 DDR3~\cite{micron-datasheet-2}, since each DIMM includes 16 and 2 DRAM chips for data storage and ECC, respectively.

\textbf{DRAM Thermal Testbed on a Server.}
To perform the experiments under controlled temperatures, we implement a temperature-controlled testbed using heating elements~\cite{DBLP:journals/dt/JungMRWW17} for DRAMs on a server. Figure~\ref{fig:thermal:a} shows the X-Gene2 board with four DIMMs fitted with our custom adapters. Each adapter consists of a resistive element, with thermally conductive tape transferring the heat of the element to all the chips in a DIMM in a uniform way, and a thermocouple to measure the temperature. The temperature of each element is controlled by a controller board, as shown in Figure~\ref{fig:thermal:b}, which contains a Raspberry Pi 3~\cite{raspberry} and four closed-loop PID controllers~\cite{ir33}.

\subsection{DRAM Parameters and Error Accounting}
\label{section:experimental:analysis}
The X-Gene2 provides access to a separate light-weight intelligent processor (SLIMpro), which is a special management core that is used to boot the system and provide access to the on-board sensors to measure the temperature and the power of the SoC and DRAM. The SLIMpro also reports all memory errors corrected or detected by SECDED ECC to the Linux kernel, providing information about the DIMM, bank, rank, row and column in which the error occurred. 
Finally, SLIMpro allows the configuration of the parameters of the MCUs, such as \trefp and \vdd. Specifically, \trefp may be changed from the nominal $64~ms$ to $2.283~s$, which is the maximum on the X-Gene2 server. The server runs a fully-fledged OS based on CentOS 7 with the default Linux kernel 4.3.0 for ARMv8 and support for 64KB pages.
\subsection{Benchmarks}\label{section:experimental:benchmarks}
In our study, we use Rodinia and Parsec benchmark suites, specifically the \emph{backprop}, \emph{nw}, \emph{srad}, \emph{kmeans} and \emph{fmm} benchmarks, which represent a variety of compute-intensive algorithms~\cite{Che:2009:RBS:1678998.1680782,Bienia:2008:PBS:1454115.1454128}.
To evaluate how parallelism and processing power affect the characterization, we run these benchmarks with 1 and 8 threads. To investigate the effect of popular caching and analytics workloads on DRAM reliability, we run the \emph{memcached} benchmark~\cite{palit-demystifying-cloud-benchmarking}, the pagerank algorithm (\emph{pagerank}), the betweenness centrality algorithm (\emph{bc}) and the breadth-first search algorithm (\emph{bfs})~\cite{Shun:2013:LLG:2517327.2442530,Sun:2017:GAL:3079079.3079097}.
Finally, we run each benchmark allocating 8 GB of data to exclude the effect of the data size factor on DRAM errors.

\section{DRAM characterization and data collection}
\label{sec:evaluation}

In this section, we characterize DRAM error behavior when running real workloads under lowered \vdd, different levels of \trefp and
the selected DRAM temperature range.

\textbf{Temperature.} We characterize DRAM at three temperature levels: $50^{\circ}C$, $60^{\circ}C$ and $70^{\circ}C$. We use this temperature range to follow previous studies~\cite{Liu:2013:ESD:2485922.2485928} and the DIMM specification~\cite{micron-datasheet}, in which the vendor reports the maximum operating temperature of $70^{\circ}C$. Note that this temperature range is common for dense server environments\cite{server_temperature,Liu:2012:RRI:2337159.2337161,8474184}.

\textbf{DRAM Circuit Parameters.} 
We experimentally determine the lowest operating DRAM \vdd as $1.428~V$, after which the circuitry of the DRAM is likely to stop working. 
We execute all the benchmarks with the memory operating under the minimum \vdd ($1.428~V$) discovered in our experiments; however, the benchmarks have not manifested errors for DRAM operating at $50^{\circ}C$. Moreover, we discover only a few CEs by running benchmarks at $60^{\circ}C$ and $70^{\circ}C$. Thus, reducing \vdd from the nominal $1.5~V$ down to $1.428~V$ (or by 5\%) has a negligible effect on DRAM reliability.

The maximum power gain is achieved when both \trefi and \vdd are scaled. To achieve this gain, in the rest of this paper, we set the minimum \vdd($1.428~V$) and run all the benchmarks under different \trefp.

\subsection{Correctable errors}
\label{subsec:correctable}

In our experiments with all the benchmarks for DRAM operating under scaled \trefp and \vdd, we encounter only CEs at $50^{\circ}C$ and $60^{\circ}C$, and no UEs or SDCs.

Previously, it was discovered that the memory cell leakage may change over time due to a phenomenon called variable retention time ($VRT$)~\cite{307481}. As a result, DRAM error behavior may vary across runs of the same application, and thus, it is essential to run each application several times until a target DRAM error metric converges to a specific value. To this end, we run each application for 2 hours with DRAM operating under the maximum \trefp (i.e. $2.283~s$) and lowered \vdd ($1.428~V$) at $50^{\circ}C$. Figure~\ref{fig:wer_in_time} shows how the rate ($WER$) of single-bit errors detected in 64-bit words changes over time for each benchmark. Note that labels with abbreviation \emph{(par)} correspond to the parallel version of the compute-intensive benchmarks. We see that after 2-hour runs $WER$ achieves a certain value for each benchmark: the average change in the $WER$ for the last 10 minutes of each experiment does not exceed 3~\% at $50^{\circ}C$. We observe the same results for DRAM operating at $60^{\circ}C$. These observations imply that 120 minutes is sufficient for identifying the vast majority of error-prone memory locations and characterize DRAM behavior when running a specific benchmark. 
\begin{figure*}[ht]
        \begin{minipage}[t]{0.33\textwidth}
                \centering
                \includegraphics[width=\textwidth,keepaspectratio]{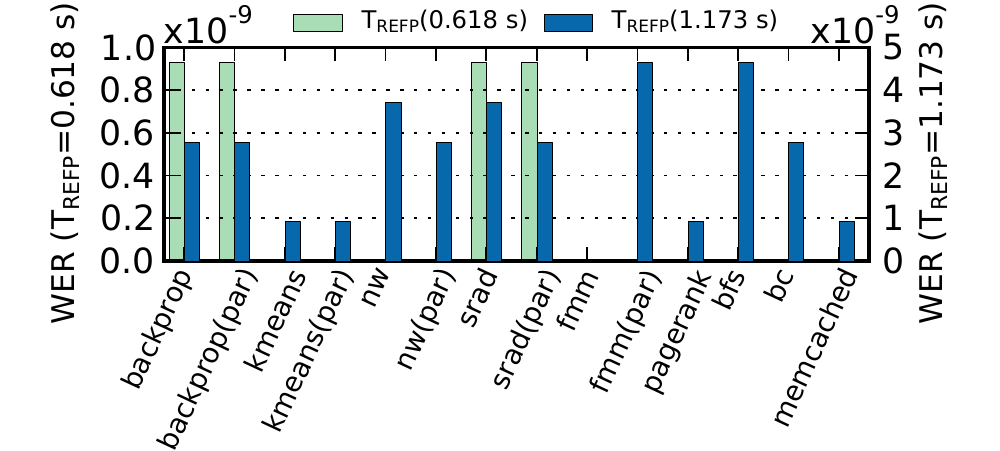}
                 (a) $50^{\circ}C$
        \end{minipage}%
        \begin{minipage}[t]{0.33\textwidth}
                \centering
                \includegraphics[width=\textwidth,keepaspectratio]{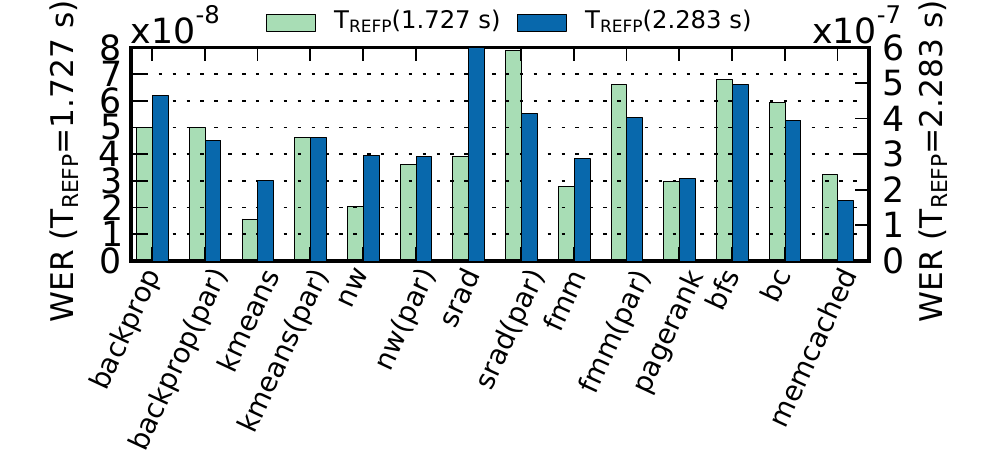}
                 (b) $50^{\circ}C$
        \end{minipage}
        \begin{minipage}[t]{0.33\textwidth}
                \centering
                \includegraphics[width=\textwidth,keepaspectratio]{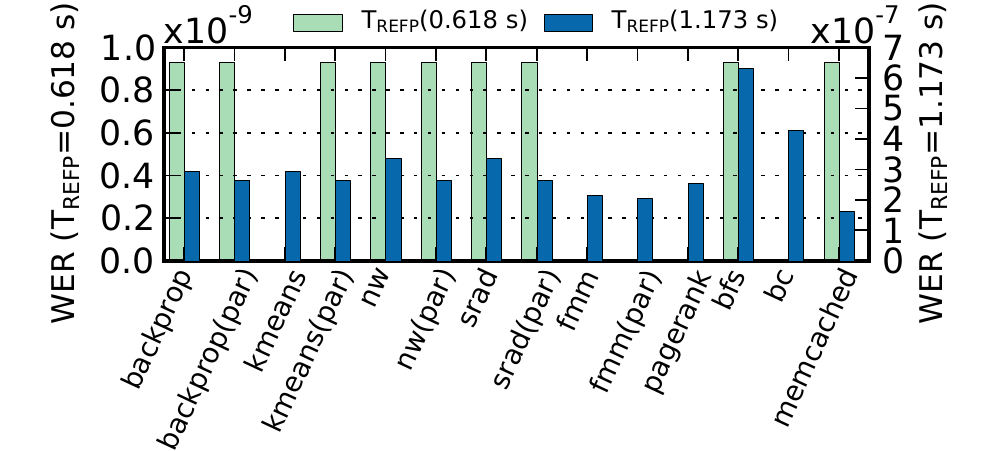}
                 (c) $60^{\circ}C$
        \end{minipage}
        \\
        \begin{minipage}[t]{0.33\textwidth}
                \centering
                \includegraphics[width=\textwidth,keepaspectratio]{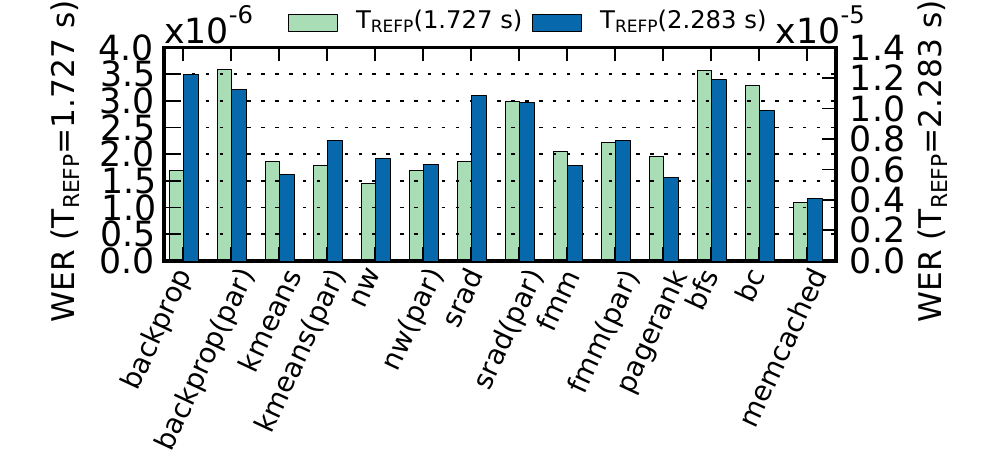}
                 (d) $60^{\circ}C$
        \end{minipage}
        \begin{minipage}[t]{0.33\textwidth}
                \centering
                \includegraphics[width=\textwidth,keepaspectratio]{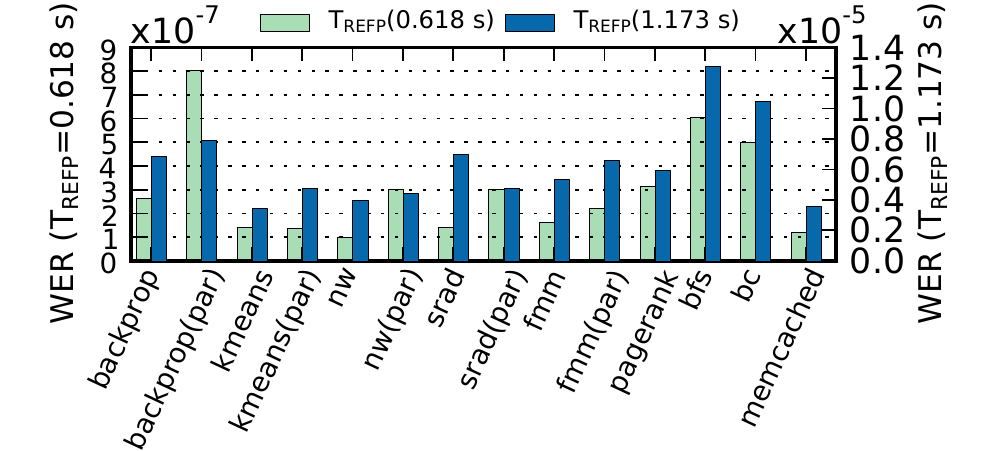}
                 (e) $70^{\circ}C$
        \end{minipage} 
        \begin{minipage}[t]{0.33\textwidth}
                \centering
                \includegraphics[width=0.55\textwidth,keepaspectratio]{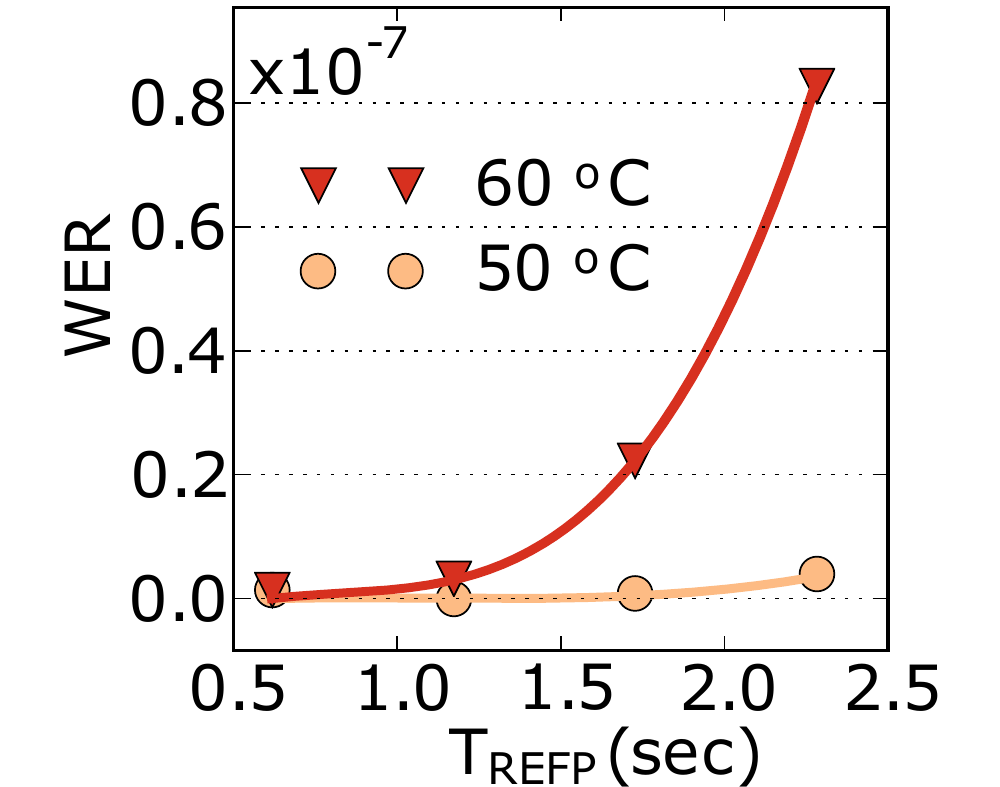}\\
                 (f) 
        \end{minipage} 
\caption{$WER$ for DRAM operating under $0.618~s$, $1.173~s$, $1.727~s$, $2.283~s$ at $50^{\circ}C$(a,b), $60^{\circ}C$(c,d) and $70^{\circ}C$(e). The $WER$ averaged over all benchmarks for DRAM operating at $50^{\circ}C$ and $60^{\circ}C$(f)}
\label{fig:wer}
\end{figure*}
\begin{figure*}[h]
        \centering
        \includegraphics[width=\textwidth]{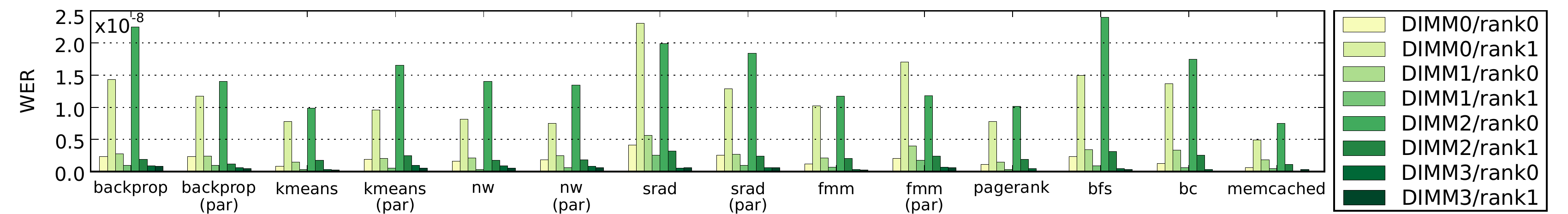}
        \caption{$WER$ per DIMM/rank obtained  for DRAM operating under $2.283~s$ \trefp at $50^{\circ}C$.}
        \label{msr_wer_per_dimm}
\end{figure*}

\begin{figure*}[ht]
        \begin{minipage}[t]{0.69\textwidth}
        \centering
        \includegraphics[width=\textwidth]{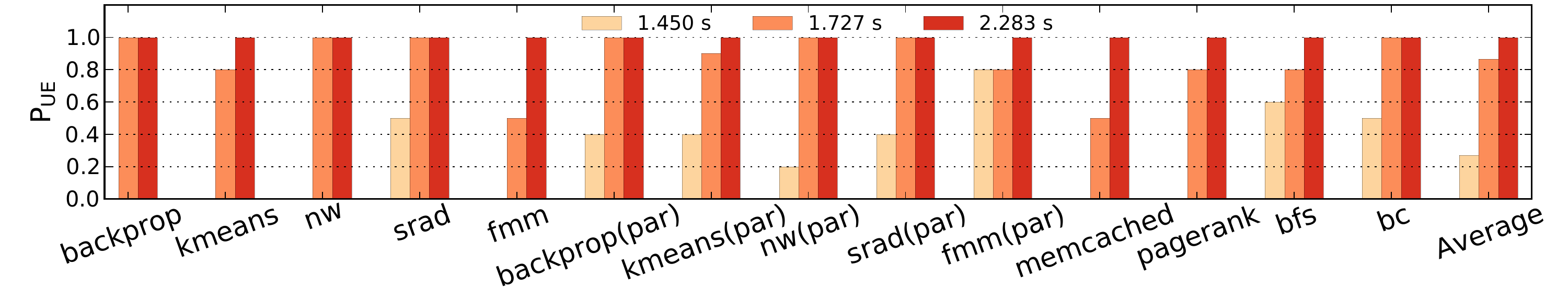}
        (a)
        \end{minipage} 
        \begin{minipage}[t]{0.3\textwidth}
        \centering
        \includegraphics[width=\textwidth, keepaspectratio]{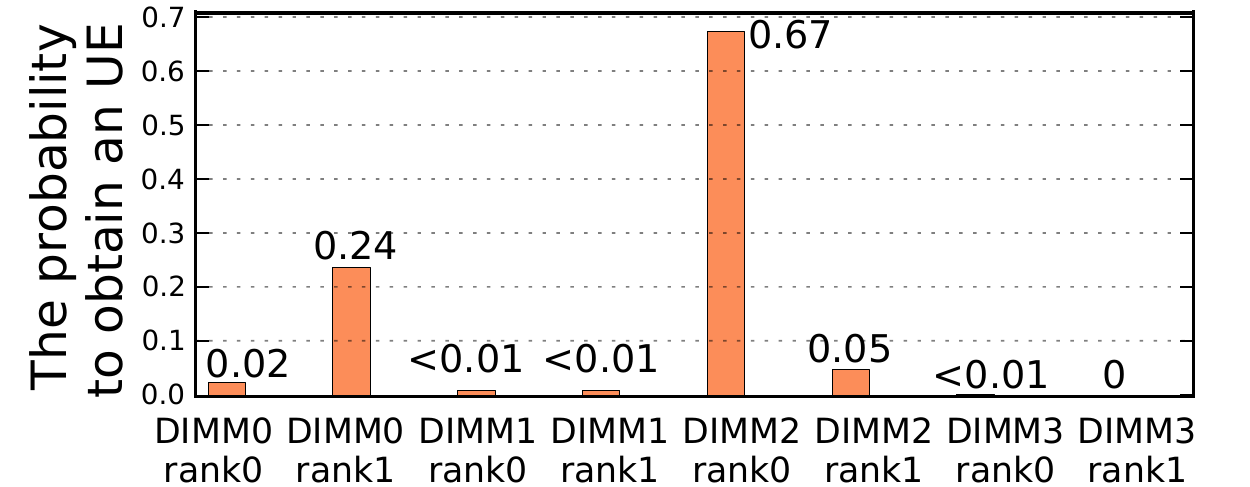}
        (b)
        \end{minipage} 
\caption{a) $P_{UE}$ and b) The probability to obtain an UE on a specific DIMM/rank when DRAM operates under $1.450~s$, $1.727~s$ and $2.283~s$~\trefp at $70^{\circ}C$.}
\label{pfail}
\end{figure*}

\begin{table}
\footnotesize
\centering
    \begin{tabular}{ | c | c | c | c | c | c |}
    \hline
      & nw & srad & backprop & kmeans & fmm \\ \hline
    1 thread & 10.93 & 2.82 & 1.61 & 0.17 &  8.88\\ \hline
    8 threads &  4.06 & 1.89 & 1.10 & 0.50 & 2.41 \\ \hline
    \end{tabular}
    \begin{tabular}{ | c | c | c | c | c | }
    \hline
      & memcached & pagerank & bfs & bc \\ \hline
    8 threads &  0.09 & 0.48 & 0.61 & 0.56 \\ \hline
    \end{tabular}

    \caption{The average DRAM reuse time.}
    \label{reuse_time}
\end{table}

\textbf{WER:} Further, we investigate how $WER$ varies across benchmarks when DRAM operates under different \trefp at $50^{\circ}C$ and $60^{\circ}C$.  We run benchmarks for DRAM operating under $0.618~s$, $1.173~s$, $1.727~s$, $2.283~s$ \trefp and lowered \vdd. Figure~\ref{fig:wer} illustrates how $WER$ changes with scaling \trefp at $50^{\circ}C$ and $60^{\circ}C$. Our first observation is that $WER$ varies across benchmarks significantly; for example, the difference achieves almost $8\times$ for \emph{memcached} and \emph{backprop(par)} when DRAM operates under $0.618~s$ at $70^{\circ}C$. Our second observation is that the benchmark that incurs the highest $WER$ may change with the DRAM temperature and \trefp; for example, the highest $WER$ is obtained for \emph{srad} when DRAM operates under $1.173~s$ at  $50^{\circ}C$, while for DRAM operating under the same \trefp at $70^{\circ}C$ the highest $WER$ is observed for the \emph{bfs} benchmark. This indicates that DRAM operational and environmental parameters may change the ratio of $WER$ between different workloads which is hard to capture with an analytical model. Our third observation is that $WER$ grows exponentially with \trefp (see Figure ~\ref{fig:wer}f).
Finally, we see that the $WER$ incurred by the parallel version of some benchmarks differs from the $WER$ obtained for the single-threaded version of these benchmarks. For example, the $WER$ measured for \textit{backprop} is almost 30~\% greater than the $WER$ obtained for \textit{backprop(par)} when DRAM operates under $2.283~s$ \trefp at $50^{\circ}C$ and $60^{\circ}C$. The same difference is also observed in the case of the \textit{srad} benchmark. Importantly, parallel and single-threaded versions of the same workload have different memory access scenarios, but a similar data pattern. Thus, these observations imply that the memory access pattern of a running program may also significantly affect DRAM error behavior.

To investigate the difference in $WER$ for parallel and single-threaded benchmarks, we calculate $T_{reuse}$ for each workload, as shown in Table~\ref{reuse_time}. We see that the $T_{reuse}$ of the parallel \textit{backprop} and \textit{srad} is less than the $T_{reuse}$ estimated for the single-threaded version of \textit{backprop} and \textit{srad}, respectively. As follows, in the case of \textit{backprop} and \textit{srad}, the parallel benchmarks implicitly refresh data in the memory more frequently than the single-threaded benchmarks do by generating more accesses to the same regions of memory per cycle. As a result, we observe a low error rate for these parallel benchmarks. Nonetheless, in the case of \textit{kmeans}, the parallel version has a higher $T_{reuse}$ ($0.50~s$) than do the serial version ($0.17~s$) due to a better data locality in caches obtained for the parallel \textit{kmeans}. Respectively, the parallel version generates fewer references to the same memory per cycle than does the single-threaded version, resulting in a higher $T_{reuse}$ and therefore a higher $WER$. Lastly, \textit{memcached} incurs the lowest $WER$ and has the lowest $T_{reuse}$ for DRAM operating under different \trefp and temperatures among all workloads at the same time, which confirms that there is a correlation between $T_{reuse}$ and DRAM error behavior.

To investigate how $WER$ varies across different DIMMs and ranks, we grouped all the collected errors by a source DIMM/rank. Figure~\ref{msr_wer_per_dimm} shows $WER$ measured on different DIMMs and ranks when DRAM operates under $2.283~s$ \trefp at $50^{\circ}C$. We see that $WER$ varies across DIMMs/ranks by up to 188x; in particular, $WER$ incurred by the \emph{bc} benchmark on DIMM2/rank0 and DIMM3/rank1 are  $1.75\times10^{-7}$ and $9.31\times10^{-10}$, respectively. Therefore, to enable accurate DRAM error predictions, a model should take into consideration the error behavior of a specific DIMM. 

\subsection{Uncorrectable Errors and System Crashes}
In our experiments with DRAM operating at $50^{\circ}C$ and $60^{\circ}C$, we have discovered no Silent Data Corruptions (SDCs) or uncorrectable errors (UEs). However, we encounter UEs and system crashes when raising the DRAM temperature to $70^{\circ}C$ and scaling \trefp up to $1.45~s$ under lowered \vdd. Note that in our framework, any UE triggered by the Linux kernel or a user-level program, once detected by ECC, will result in a system crash.

Figure~\ref{pfail}a shows $P_{UE}$, the likelihood to observe an UE, measured across all benchmarks for DRAM operating under $1.450~s$, $1.727~s$, $2.283~s$ \trefp and lowered \vdd at $70^{\circ}C$. To estimate this probability, we repeat each 2-hour experiment with a specific benchmark 10 times. We see that $P_{UE}$ varies significantly across benchmarks for DRAM operating under $1.450~s$ \trefp; it achieves $0.8$ for \textit{fmm(par)}, whereas it equals to $0$ for \textit{memcached} and \textit{pagerank}. We also observe that $P_{UE}$ is greater than $0$ only for the parallel compute-intensive benchmarks, while it is $0$ for all the single-threaded benchmarks except for \textit{srad}. 
The $P_{UE}$ averaged over all benchmarks for DRAM operating under $1.450~s$ \trefp is below $0.4$. However, when we increase \trefp up to $1.727~s$, then the likelihood of crashing averaged over benchmarks grows by $2.15\times$ (see Figure~\ref{pfail}). Moreover, for DRAM operating under this \trefp, there is no benchmark with $P_{UE}=0$. Finally, all the benchmarks trigger UEs in 100\% of the experiments when we use the maximum \trefp($2.283~s$) at $70^{\circ}C$. These results show that \trefp and the DRAM temperature have a dominant effect on the likelihood of an UE.

Figure~\ref{pfail}b depicts the probability to obtain an UE on a specific DIMM/rank when ECC detects an UE. We see that the vast majority of UEs are triggered by DIMM0/rank1 and DIMM2/rank0, while DIMM3/rank1 do not trigger UEs at all. Thus, DRAM reliability varies significantly from DIMM-to-DIMM not only in terms of $WER$ but also the probability to obtain an UE.
Importantly, we have discovered no SDCs when running experiments under different \trefp at $50^{\circ}C$, $60^{\circ}C$, and $70^{\circ}C$.

\section{Accuracy evaluation of ML models}
\label{sec:model_evaluation}
In this section, we present the results of the feature selection process and accuracy evaluation of ML models.

\subsection{Feature selection}
The accuracy of an ML model strongly depends on the set of features used for training of the model. If the model is trained using the set of features that are not correlated with a metric that we target to predict, then the model may overestimate the significance of some features~\cite{Domingos:2012:FUT:2347736.2347755}. As a result, a low prediction accuracy will be obtained for this model. To identify those features that may affect DRAM reliability, we extract 249 program features, including $T_{reuse}$ (the average memory reuse time) and $H_{DP}$ (the data entropy, see Section~\ref{sec:modeling}), for each benchmark, and correlate them with both $WER$ and $P_{UE}$ metrics.

\textbf{WER:} We build the correlation of $WER$ and program features using the combined measurements taken under different levels of \trefp ($0.618~s$, $1.173~s$, $1.727~s$, $2.283~s$) at $50^{\circ}C$, $60^{\circ}C$ and $70^{\circ}C$, where we observe no UEs or system crashes. To identify and quantify any dependency between program features and the DRAM error metrics formally, we use the Spearman's rank correlation coefficient ($r_{s}$). This correlation coefficient allows us to detect both linear and non-linear relationships~\cite{8703893}. Coefficient values lie in a range $[-1,+1]$ in which -1 or +1 occurs when there is a perfect monotonic relationship between two variables. 
\begin{wrapfigure}{L}{0.6\columnwidth}
\centering
\includegraphics[height=1.7in]{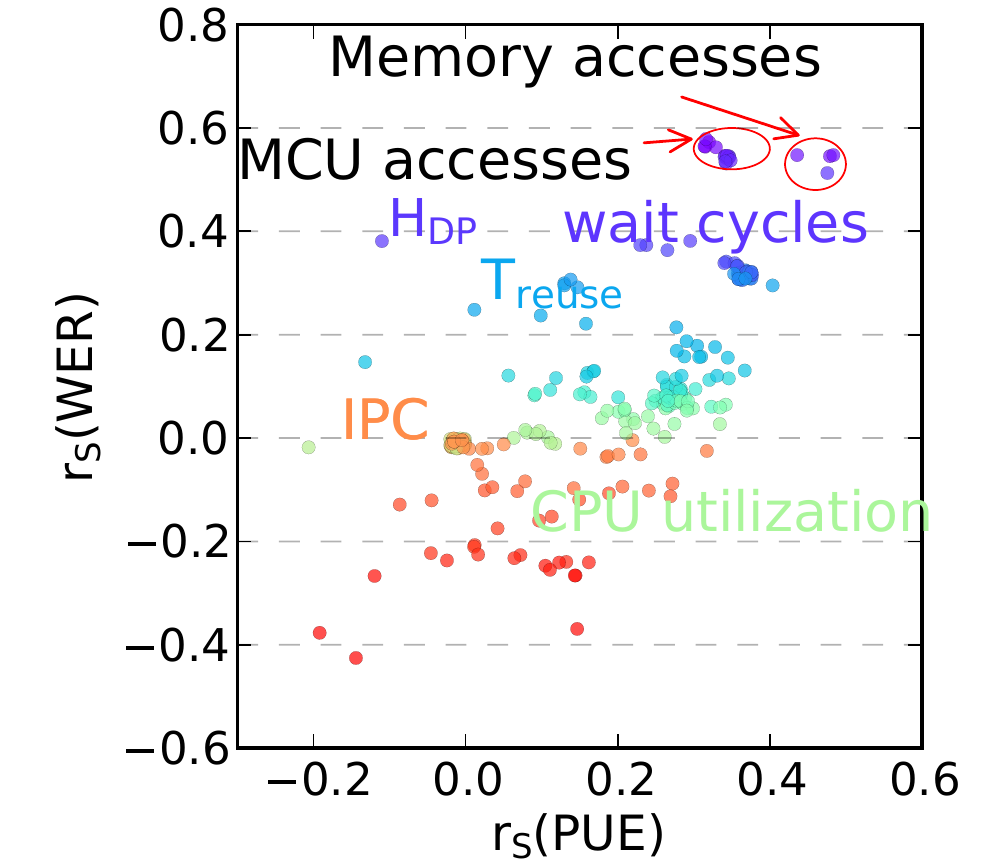}
\caption{ $r_{s}$ for 249 program features and $WER$ and $P_{UE}$.}
\label{fig:correlation}         
\end{wrapfigure}

\remove{
\begin{table}[h]
  \begin{minipage}[t]{0.45\linewidth}
  \centering
  \includegraphics[height=1.6in]{figures/correlation_ber_pcrash.pdf}
  \caption{ $r_{s}$ for 249 program features and $WER$ and $P_{UE}$}
  \label{fig:correlation} 
  \end{minipage}
  \begin{minipage}[b]{0.45\linewidth}
    \footnotesize
    \centering
    \begin{tabular}{ | c | c | }
    \hline
    \textbf{Set} & \textbf{Parameters}\\ \hline
    1 & $TEMP_{DRAM}$, \trefp, \\
      & wait cycles, \\
      & memory accesses \\ 
      & $H_{DP}$, $T_{reuse}$ \\ 
    \hline
    2 & $TEMP_{DRAM}$, \trefp, \\
      & wait cycles, \\
      & memory accesses \\ 
    \hline
    3 & $TEMP_{DRAM}$, \trefp,\\
      & all program features\\ 
    \hline
    \end{tabular}
    
    \caption{Input feature sets used for $SVM$}
    \label{input_parameters}
  \end{minipage}
\end{table}
}
Figure~\ref{fig:correlation} shows the correlation coefficients for 249 program features and $WER$ on the Y-axis, whereas the correlation coefficients for these features and $P_{UE}$ are shown on the X-axis.
We see that the number of memory accesses per cycle is highly correlated with $WER$, as $r_s$ is above 0.57, indicating a positive direction of the correlation; in other words, $WER$ grows with the number of accesses per cycle. We also observe that the group of performance indicators that reflects the number of issued memory read and write commands per cycle in different MCUs is also highly correlated with $WER$. However, the number of such commands is determined by the number of memory read and write instructions executed by the processor per cycle. 

Another inherent program feature that is strongly correlated with $WER$ is \textit{wait cycles} ($r_s$ is 0.4). This feature reflects the ratio of the number of cycles spent on waiting for data to the total number of program cycles. Nonetheless, \textit{wait cycles} is implicitly determined by the number of memory accesses per clock cycle, as it encapsulates idle cycles due to memory access stalls which explains its correlation with $WER$.

We attribute the correlation of the memory access rate and $WER$ to disturbance errors induced by the cell-to-cell interference~\cite{5727538,1029773}. Previously, it was shown that, if a row is accessed many times, then some cells from neighbouring rows may leak charge quickly~\cite{6853210}. Thus, by accessing the memory with a high rate, we increase the probability of the interference errors for DRAM operating under scaled \trefp and \vdd. By contrast, under a higher memory access rate, each cell may be implicitly refreshed more frequently, resulting in a lower $WER$. However, this effect occurs only for those benchmarks in which $T_{reuse}<T_{REFP}$. Therefore, a high memory access rate may have negative or positive effects on DRAM reliability, which depends on $T_{reuse}$ and \trefp. Notably, $T_{reuse}$ is greater than the maximum \trefp ($2.283~s$) available on our platform for almost 30~\% of the benchmarks. Thereby, $T_{reuse}$ does not have any effect on DRAM error behavior in these benchmarks. This lack of an effect explains why $T_{reuse}$ ($r_s$ is 0.23) is less correlated with $WER$ than the rate of memory accesses.

Our experiments show that $H_{DP}$, which reflect the data pattern of a running application, is also correlated with $WER$ as the $r_s$ is 0.39, see Figure~\ref{fig:correlation}. Although it is higher than the $r_s$ obtained for $T_{reuse}$, it is by 51~\% lower than the $r_s$ observed for the memory access rate.

\textbf{The probability of an UE:} 
Similar to $WER$, we discover a correlation between $P_{UE}$ and the memory access rate, the number of issued memory read and write commands per cycle in different MCUs, $H_{DP}$, and \textit{wait cycles}. However, the level of this correlation is lower than in the case of $WER$; for example, the $r_s$ for the memory access rate and $P_{UE}$ is 0.43, which is 35~\% less than the same $r_s$ for $WER$. It is noteworthy that unlike previous studies, which have indicated a strong impact of  $T_{reuse}$ or $H_{DP}$ \cite{Khan:2014:EEM:2637364.2592000,8613828}, we obtain the highest $r_s$ for the memory access rate among all the program features when correlating it with $WER$ and $P_{UE}$ metrics.

\textbf{Implication:} \emph{Thus, our study indicates that the memory access rate has a major effect on DRAM reliability, which is stronger than the effect of the content data stored in DRAM and the average DRAM reuse time.}

\subsection{Accuracy evaluation}
\begin{table}
\footnotesize
\centering
    \begin{tabular}{ | c | c | }
    \hline
    \textbf{Input set} & \textbf{Parameters}\\ \hline
    1 & $TEMP_{DRAM}$, \trefp, wait cycles \\ 
      & memory accesses, $H_{DP}$, $T_{reuse}$ \\ 
    \hline
    2 & $TEMP_{DRAM}$, \trefp, wait cycles \\ 
      & memory accesses \\ 
    \hline
    3 & $TEMP_{DRAM}$, \trefp, all program features\\ 
    \hline
    \end{tabular}
    
    \caption{Input feature sets used for training}
    \label{input_parameters}
\end{table}

\textbf{WER:} We start our evaluation campaign by applying SVM, KNN and RDF models to predict $WER$ using 3 different input sets of parameters (see Table~\ref{input_parameters}), which consist of different combinations of program features, \trefp and the DRAM temperature ($TEMP_{DRAM}$). Note that we investigate different input sets, as it is known that the accuracy of an ML model depends on the input parameters that are chosen for training~\cite{Dougherty:1995:SUD:3091622.3091646}. We build the first two input sets using the program features that are strongly correlated with DRAM error behavior. In the third set of input parameters, we include all the collected program features, to investigate the model accuracy when all the available parameters are provided to the model. 

\begin{figure*}[h]
        \begin{minipage}[t]{0.33\textwidth}
                \centering
                \includegraphics[width=\textwidth,keepaspectratio]{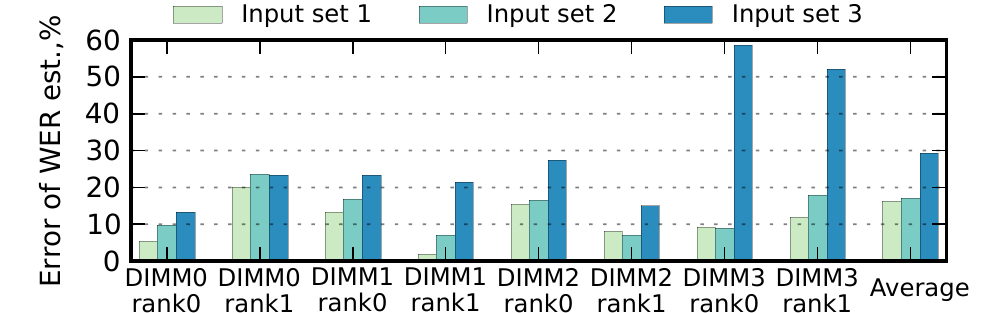}
                 (a) SVM
        \end{minipage}%
        \begin{minipage}[t]{0.33\textwidth}
                \centering
                \includegraphics[width=\textwidth,keepaspectratio]{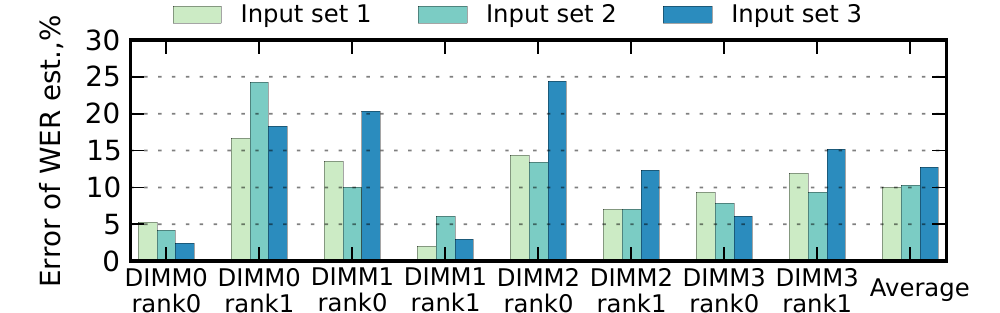}
                 (b) KNN
        \end{minipage}
        \begin{minipage}[t]{0.33\textwidth}
                \centering
                \includegraphics[width=\textwidth,keepaspectratio]{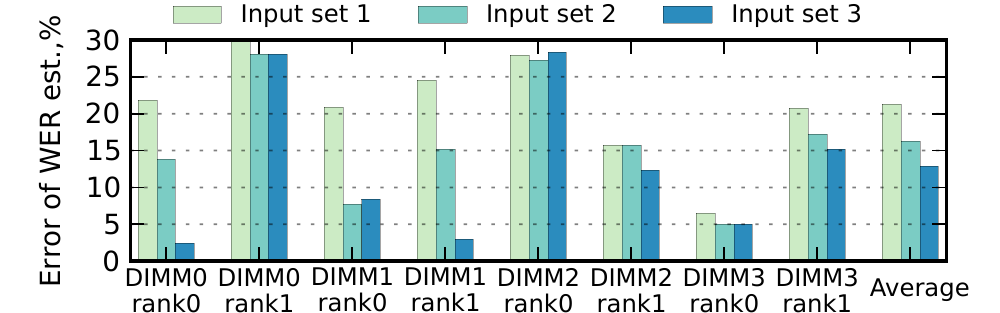}
                 (c) RDF
        \end{minipage}
        \\
        
        \begin{minipage}[t]{0.33\textwidth}
                \centering
                \includegraphics[width=\textwidth,keepaspectratio]{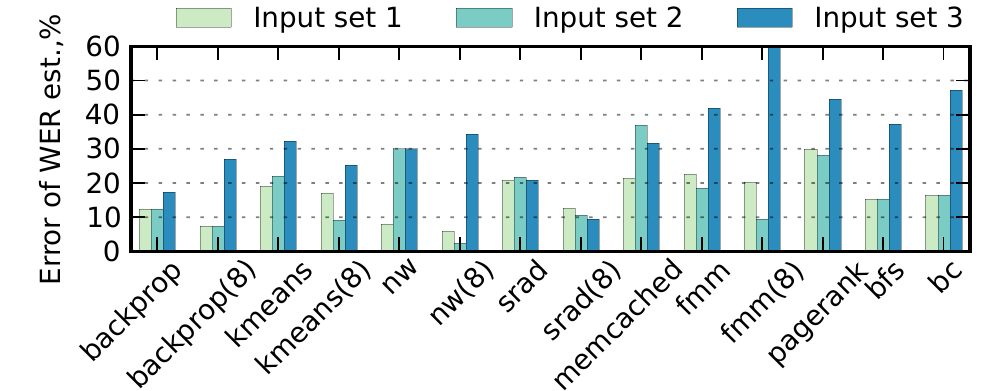}
                 (d) SVM
        \end{minipage}%
        \begin{minipage}[t]{0.33\textwidth}
                \centering
                \includegraphics[width=\textwidth,keepaspectratio]{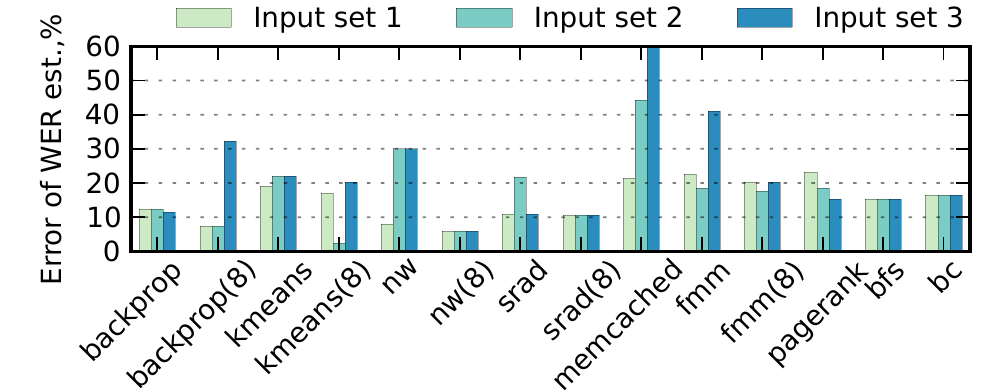}
                 (e) KNN
        \end{minipage}
        \begin{minipage}[t]{0.33\textwidth}
                \centering
                \includegraphics[width=\textwidth,keepaspectratio]{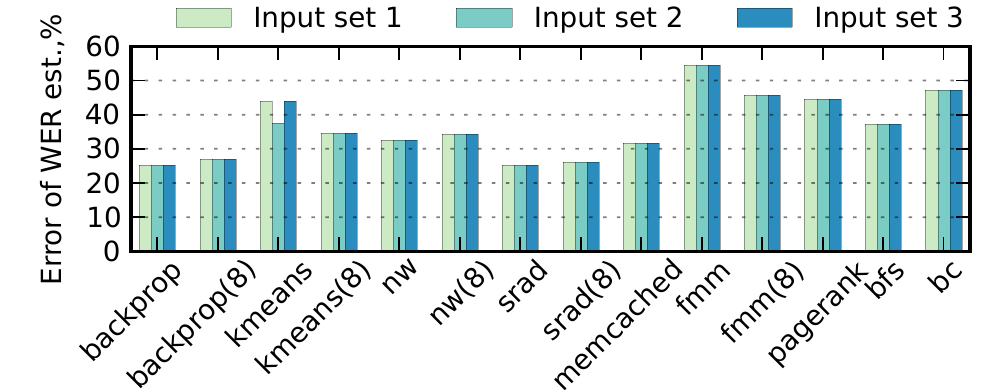}
                 (f) RDF
        \end{minipage}
\remove{
        \\
        \begin{minipage}[t]{0.33\textwidth}
                \centering
                \includegraphics[width=\textwidth,keepaspectratio]{figures/pcrash_per_application.pdf}
                 (g) SVM
        \end{minipage}%
        \begin{minipage}[t]{0.33\textwidth}
                \centering
                \includegraphics[width=\textwidth,keepaspectratio]{figures/pcrash_per_application_knc.pdf}
                 (h) KNN
        \end{minipage}
        \begin{minipage}[t]{0.33\textwidth}
                \centering
                \includegraphics[width=\textwidth,keepaspectratio]{figures/pcrash_per_application_rfc.pdf}
                 (i) RDF
        \end{minipage}
        }
\caption{The average error of $WER$ estimates per DIMM/rank: a) SVM, b) KNN and c) RDF. The average error of $WER$ estimates per application: d) SVM, e) KNN and f) RDF. 
}
\label{fig:model_accuracy}
\end{figure*}

Figure ~\ref{fig:model_accuracy} (a,b,c) shows the mean percentage error ($MPE$) of $WER$ estimates provided by SVM, KNN and RDF per DIMM/rank for all three sets of input parameters. We see that the minimum error of $WER$ estimates averaged over all the DIMMs and ranks is achieved when we use the first set of input parameters for SVM (16.3~\%) and KNN (10.1~\%), while the average error incurred by the second input set for SVM and KNN are 17.0~\% and 10.2~\%, respectively. Thereby, by adding $H_{DP}$ and  $T_{reuse}$ to the input parameter set, we only slightly increase the accuracy of the two models. This implies that the memory access rate has the strongest impact on DRAM error behavior among all the program features, which is consistent with the results of the feature selection process. 

Notably, if we train SVM and KNN using all the collected program features for each workload, then the average $MPE$ grows up to 29.3~\% (SVM) and 12.3~\% (KNN). We explain this by overfitting of the model which happens when we train it using all the available program features, including those that do not affect DRAM reliability. 
In other words, the models may overestimate the significance of some features when we train the model using all the features, which results in a low prediction accuracy obtained for the third set~\cite{Domingos:2012:FUT:2347736.2347755}.

Interestingly, in contrast to SVM and KNN, RDF provides the lowest accuracy of $WER$ estimates (the error is 21.4~\%) when the first input set is used. Moreover, this model demonstrates the highest accuracy (the error is 12.9~\%) when all the available program features are used for training and testing. Nonetheless, this accuracy is less than the best accuracy achieved by KNN when the first input set is used. Furthermore, the maximum error of $WER$ estimated per application is about 55~\% when we use the third input set for the RDF model, see Figure~\ref{fig:model_accuracy}f (the \emph{fmm} benchmark). Meanwhile, the average error of $WER$ estimates provided by SVM and KNN per application do not exceed 30~\% and 24~\%, correspondingly, when we use the first input set. Thus, we may conclude that RDF has the lowest accuracy among the considered models when predicting $WER$.

\begin{figure}[h]
        \begin{minipage}[t]{0.49\columnwidth}
                \centering
                \includegraphics[width=\textwidth,keepaspectratio]{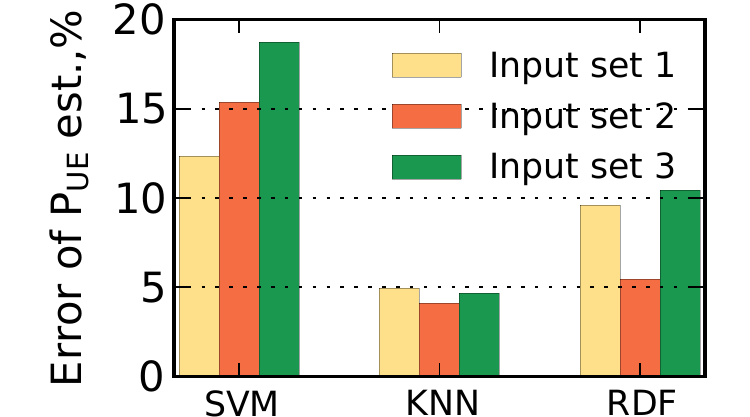}
                \caption{The error of $P_{UE}$ estimates averaged over applications and DIMMs.}
                \label{fig:pue_accuracy} 
        \end{minipage}%
        \hfill%
        \begin{minipage}[t]{0.49\columnwidth}
                \centering                \includegraphics[width=\textwidth,keepaspectratio]{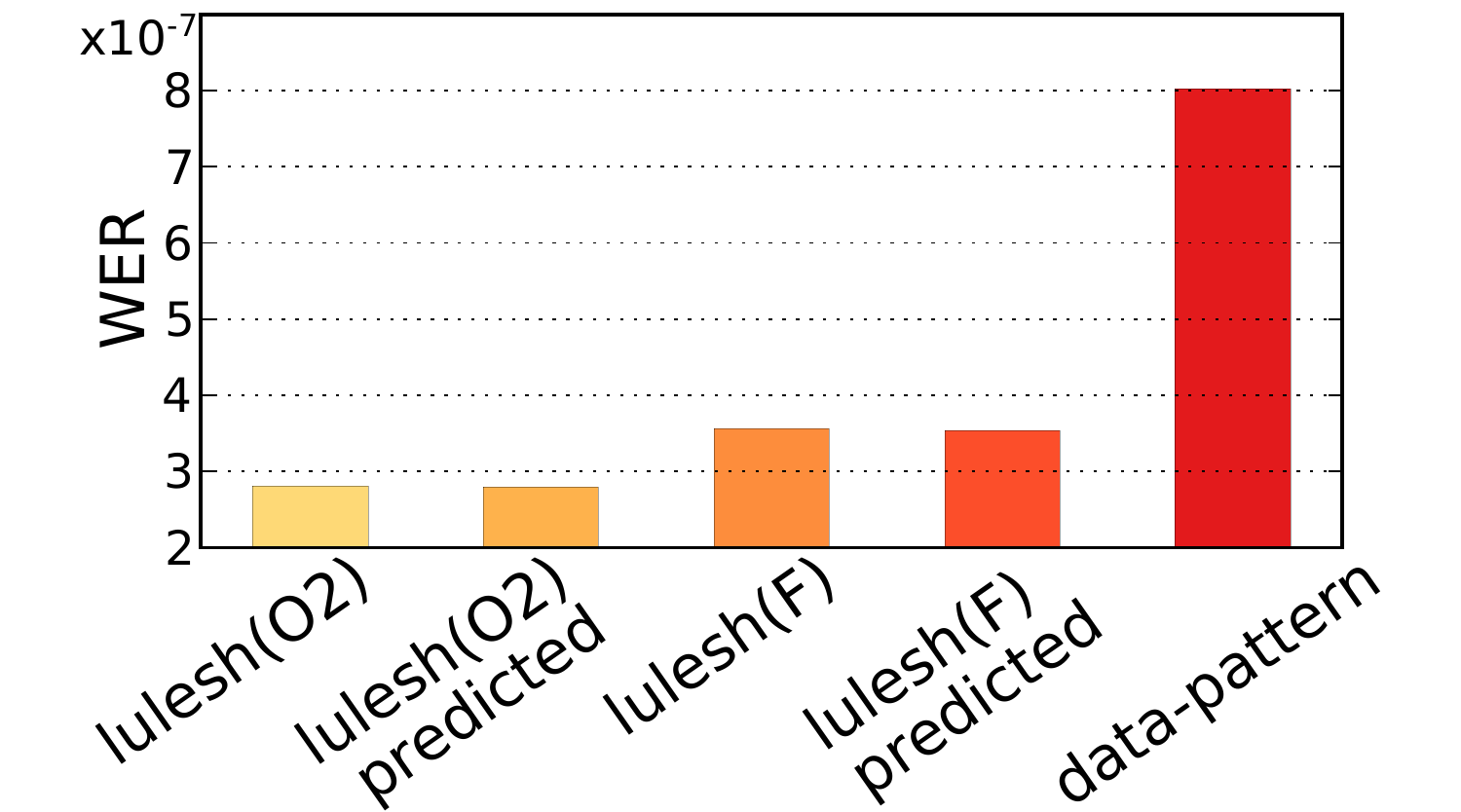}
                \caption{The measured and predicted $WER$ for \emph{lulesh} and the \emph{random} micro-benchmark (\trefp is 0.618~s, $70^{\circ}C$).}
                \label{fig:lulesh_predict} 
        \end{minipage}
\end{figure}

\textbf{The probability of an UE:} Figure~\ref{fig:pue_accuracy} depicts the mean percentage error of $P_{UE}$ estimates averaged over all benchmarks and DIMMs. Similar to our experiments with $WER$, we see that the first set incurs the lowest error (12.3~\%) when we use SVM. While the average error obtained by this model for the second and third sets is above 15~\%. However, KNN and RDF demonstrate the lowest average error when we use the second input feature set. Notably, this error is only 4.1~\% and 5.5~\% for KNN and RDF, respectively, which is almost 3$\times$ lower than the lowest error (12.3~\%) achieved by SVM.

To conclude, our study shows that the highest accuracy of $WER$ estimates is achieved by the K-nearest neighbors algorithm (KNN) when we train it using the first input set of parameters (i.e. the memory access rate, \textit{wait cycles}, $H_{DP}$ and  $T_{reuse}$, $TEMP_{DRAM}$, \trefp and \vdd). The highest accuracy of $P_{UE}$ estimates is also demonstrated by KNN when we use the second input set, which contains only the memory access rate, \textit{wait cycles}, $TEMP_{DRAM}$ and \trefp.

\subsection{Workload-Aware Modeling vs Conventional Modeling}
Many studies have proposed to model DRAM errors for investigation either hardware design efficiency \cite{Liu:2011:FSD:1961296.1950391} or software fault tolerance ~\cite{6903603,Li:2010:REM:1855840.1855846,1347082,Li:2007:ACI:1317533.1318113}. However, all those studies use constant DRAM error rates extracted on real DRAMs  when running the data pattern micro-benchmarks~\cite{Liu:2012:RRI:2337159.2337161,7266870,6579591,DBLP:journals/dt/JungMRWW17,678551}. Our model can be used to improve those studies and proposed techniques by considering workload-aware DRAM error behavior. For example, Figure~\ref{fig:lulesh_predict} depicts the measured $WER$ over all DIMMs when DRAM is operating under $0.618~s$ \trefp at $70^{\circ}C$ for the \emph{lulesh} benchmark and a data pattern micro-benchmark that implements a random data pattern~\cite{Khan:2014:EEM:2637364.2592000}. This figure also shows the $WER$ which has been predicted by the KNN-based DRAM error behavioral model. In this experiments, we use two versions of \emph{lulesh} to illustrate the implicit effect of compiler optimizations on DRAM reliability: the benchmark compiled with $-O2$ (default optimizations) and $-F$ (aggressive optimizations). We see that the model correctly predicts the $WER$ incurred by both versions of the benchmark; the error is less than 3~\%. Such a high accuracy enables us to correctly predict the difference in $WER$ between these benchmarks, which is about 29~\%. At the same time, we see that the \emph{random} micro-benchmark incurs the $WER$ which is higher than the $WER$ obtained for \emph{lulesh} by 2.9$\times$. Thus, the conventional DRAM error modeling based on  the constant rates may be inaccurate and lead to incorrect conclusions about the effectiveness of 
applied techniques.

Moreover, the vast majority of research studies have considered only hardware-level techniques to mitigate errors for DRAM operating under scaled \cite{7266870,6579591,DBLP:journals/dt/JungMRWW17,678551}, which introduce additional power and chip area overheads. However, as we see, even compiler optimizations may implicitly affect DRAM error behavior. To systematically study the effect of compiler optimizations, it is essential to build a model, since such a study may take months or even years if it is conducted using DRAM characterization campaigns. While our models predict DRAM errors within 300 ms, which opens new avenues for research.

\section{Related Work}
\label{sec:related_work}

\textbf{Scaling of \trefp and \vdd:}
Many studies\cite{1255476,Liu:2012:RRI:2337159.2337161,Ohsawa:1998:ODR:280756.280792,7266870,Venkatesan:2006:RAPID,802898} tried to improve DRAM performance and energy efficiency by adopting a low refresh period for "weak" cells. The main idea of such an approach is to split memory cells into groups based on their retention time and relax the refresh rate for those groups where cells have small leakage. Other works~\cite{4408251,4731174,refrint} suggested to skip refresh operations for those memory segments that have been implicitly refreshed by memory accesses. Several studies~\cite{Isen:2009:EEN:1669112.1669156,6579591} proposed to extend this technique and refresh selectively only rows with valid data allocated by running applications or OS. Chang et al.~\cite{Chang:2017:URO:3107080.3084447} provided the results of their study on reduced-voltage operation in DDR3L memory devices. 
However, even though the latest study~\cite{Khan:2017:DMD:3123939.3123945} tried to capture the effect of varying data patterns on DRAM reliability when running real applications, all these studies ignored the combined effect of data and memory access patterns on DRAM errors. To the best of our knowledge, none of previous works have systematically investigated the combined impact of these patterns on memory errors on a real server. Understanding of such an impact is crucial for facilitating the co-design of software and hardware techniques to improve DRAM energy efficiency. 
Other research studies proposed various fine-grained schemes to reduce the number of refresh operations and thus improve DRAM energy efficiency\cite{Stuecheli:2010:ERT:1934902.1934983,4408251,Isen:2009:EEN:1669112.1669156,Kotra:2017:HCM:3093315.3037724,Valero:2015:RRP:2802199.2802327,8456584}. Although some of these studies utilize workload inherent features, such as the memory reuse time, they are orthogonal to our work.

\textbf{Predictive maintenance and statistical prediction of errors:}
Considerable research has been done on statistical prediction of different types of hardware faults, including DRAM errors, in supercomputers~\cite{Giurgiu:2017:PDR:3154448.3154451,Lan:2010:SDM:1786811.1787078,1633531,Sahoo:2003:CEP:956750.956799,Yu:2011:POF:2056318.2057092,7575388,DBLP:conf/dsn/NieXGPEST18,Meza:2015:RME:2859844.2859952}. The majority of these studies proposed different techniques, based either on rules~\cite{1633531} or Machine Learning ~\cite{Sahoo:2003:CEP:956750.956799}, for prediction of failures that may  happen in various hardware components using history of errors. Other research studies tried to systematically investigate factors, including workload-dependent factors, that may affect DRAMs in data centers and supercomputers \cite{Schroeder:2009:DEW:1555349.1555372,Meza:2015:RME:2859844.2859952,Sridharan:2012:SDF:2388996.2389100,6877455}.
Nonetheless, all these studies tried to predict errors for hardware operating under nominal parameters.    

Hardware error prediction becomes extremely important in production lines for identifying maintenance cycles or faulty components (\emph{predictive maintenance})~\cite{Giurgiu:2017:PDR:3154448.3154451,Schroeder:2009:DEW:1555349.1555372}
However, any study of failures for hardware operating under nominal parameters may require years~\cite{Giurgiu:2017:PDR:3154448.3154451}, while a reliability characterization of hardware that operates under relaxed parameters is much faster.
In our future research, we aim to investigate how characterization and modeling of errors for DRAM operating under relaxed parameters can be applied to identify maintenance cycles or any abnormal hardware behavior.

\section{Conclusion}
\label{sec:conclusions}
In this work, we present the results of a study on characterization and prediction of the error behavior for DRAM operating under scaled parameters within a real server. Our results indicate that the rate of single- and multi-bit errors may vary across workloads and DRAM chips by 8$\times$ and 188$\times$, respectively. We quantify the effect of inherent program features that may significantly affect DRAM errors by correlating 249 features extracted from various benchmarks with DRAM errors. We train three ML models to predict DRAM failure rates and compare the accuracy of the models using different sets of program features. We demonstrate that, with the correct choice of program features and an ML model, the word-error-rate for single-bit failures and the likelihood of a system crash triggered by uncorrectable errors can be predicted for a specific DRAM device with an average error of less than 10.5~\%.

\section*{Acknowledgment}
This work was funded by the H2020 Framework Program of the European Union through the UniServer Project (Grant Agreement 688540, http://www.uniserver2020.eu) and OpreComp project (Grant Agreement 732631, http://oprecomp.eu). We are grateful to Dr. Philip Hodgers (ECIT) for providing the thermal testbed.

\bibliographystyle{plain}
\bibliography{iiswc.bib}

\end{document}